\documentclass[12pt]{article}
\usepackage{amsmath}
\usepackage{amssymb}
\usepackage{amsthm}
\newcommand{\pa}{\partial}
\newcommand{\bm}[1]{\textnormal{\mathversion{bold}$#1$}}
\newcommand{\ul}[1]{\underline{#1}}

\newcommand{\defeq}{\stackrel{\mathrm{def}}{=}}

\newcommand{\vac}{|\mathrm{vac}\rangle}
\newcommand{\dvac}{\langle\mathrm{vac}|}

\newcommand{\Ftor}[1]{{\cal F}^{\mathrm{tor}}_{#1}}

\newcommand{\IC}{\mathbb{C}}
\newcommand{\IN}{\mathbb{N}}
\newcommand{\IR}{\mathbb{R}}
\newcommand{\IZ}{\mathbb{Z}}

\newcommand{\fraksl}{\mathfrak{sl}}
\newcommand{\frakg}{\mathfrak{g}}

\newcommand{\SL}{\mathop{{\mathrm{SL}}}}

\newcommand{\gtor}{\mathfrak{g}^{\mathrm{tor}}}
\newcommand{\slt}{\mathfrak{sl}^{\mathrm{tor}}}
\newcommand{\slh}{\widehat{\mathfrak{sl}}}

\newcommand{\ii}{\mbox{{\rm i}}}
\newcommand{\ee}{\mbox{{\rm e}}}
\newcommand{\dd}{\mbox{{\rm d}}}
\newcommand{\by}{\bar{y}}
\newcommand{\bz}{\bar{z}}
\newcommand{\OmegaTor}{\Omega^{\mathrm{tor}}}

\newtheorem{theorem}{Theorem}
\newtheorem{proposition}{Proposition}
\newtheorem{lemma}{Lemma}
\newtheorem{corollary}{Corollary}

\title{
Hierarchy of $(2+1)$-dimensional nonlinear Schr\"odinger equation, 
self-dual Yang-Mills equation, and toroidal Lie algebras}
\author{
Saburo Kakei
\footnote{Present address: 
Department of Mathematics, Rikkyo University, 
Nishi-ikebukuro 3-34-1, Toshima-ku, Tokyo 171-8501, Japan. 
E-mail: kakei@rkmath.rikkyo.ac.jp}\\
{\normalsize Department of Mathematical Sciences, }\\
{\normalsize School of Science and Engineering, Waseda University}\\
{\normalsize Ohkubo 3-8-1, Shinjyuku-ku, Tokyo 169-8555, Japan}\\[3mm]%
Takeshi Ikeda\\
{\normalsize Department of Applied Mathematics,
Okayama University of Science}\\
{\normalsize Ridaicho 1-1, Okayama 700-0005, Japan}\\
{\normalsize E-mail: ike@xmath.ous.ac.jp}\\[3mm]%
Kanehisa Takasaki\\
{\normalsize Department of Fundamental Sciences, }\\
{\normalsize Faculty of Integrated Human Studies, Kyoto University}\\
{\normalsize Yoshida, Sakyo-ku, Kyoto 606-8501, Japan}\\
{\normalsize E-mail: takasaki@math.h.kyoto-u.ac.jp}}
\date{}
\begin{document}
\begin{titlepage}
\maketitle

\begin{abstract}
The hierarchy structure associated with a $(2+1)$-dimensional 
Nonlinear Schr\"odinger equation is discussed as an extension of the 
theory of the KP hierarchy. 
Several methods to construct special solutions are given. 
The relation between the hierarchy and a 
representation of toroidal Lie algebras are established 
by using the language of free fermions.
A relation to the self-dual Yang-Mills equation is also discussed. 
\end{abstract}
\end{titlepage}
\section{Introduction}
\setcounter{equation}{0}%
There have been many studies on multi-dimensional integrable 
evolution equations. An example of such equations was given by Calogero
\cite{Cal}, which is a $(2+1)$-dimensional extension of the Korteweg-de
Vries equation, 
\begin{equation}
u_t=\frac{1}{4}u_{xxy} + uu_y + \frac{1}{2}u_x \int^x u_y \dd x.
\label{2dKdV}
\end{equation}
Bogoyavlensky \cite{Bog1} showed that there is a hierarchy of
higher-order integrable equations associated with \eqref{2dKdV}. 
In the previous paper \cite{IT}, two of the present authors generalized the
Bogoyavlensky's hierarchy based on the Sato theory of the
Kadomtsev-Petviashvili (KP) hierarchy \cite{Sato,SS,DJKM,JM,JMD,UT}, and 
discussed the relationship to toroidal Lie algebras. 
We note that the relation between integrable hierarchies and toroidal
algebras has been discussed also by Billig \cite{Bi}, Iohara, Saito and
Wakimoto \cite{ISW1,ISW2} by using vertex operator representations. 

In this paper, we shall consider a $(2+1)$-dimensional extension 
of the nonlinear Schr\"odinger (NLS) equation \cite{Bog2,Sc,St1,St2}, 
\begin{equation}
\ii u_T + u_{XY} + 2u\int^X (| u|^2)_Y \dd X = 0, 
\label{2dNLS}
\end{equation}
and the hierarchy associated with this equation. In the case $X=Y$, this 
equation is reduced to 
\begin{equation}
\label{NLS}
\ii u_{T} + u_{XX} + 2 |u|^2 u =0, 
\end{equation}
which is the celebrated Nonlinear Schr\"odinger (NLS) equation. 
Equation \eqref{2dNLS} is related to the self-dual Yang-Mills (SDYM)
equation, and has been studied by several researchers from various
viewpoints: Lax pairs \cite{Bog2,Sc,St1}, Hirota bilinear method
\cite{SOM,St2}, twistor approach \cite{St2}, Painlev\'e analysis
\cite{JB}, and so on. 
Strachan \cite{St2} pointed out that \eqref{2dNLS} is transformed to 
Hirota-type equations, 
\begin{equation}
\label{Bi2dNLS}
(\ii D_T + D_X D_Y)G\cdot F =0, \qquad D_X^2 F\cdot F=2G\bar{G}, 
\end{equation}
with the transformation $u=G/F$. 
Here we have used the Hirota's $D$-operators, 
\begin{equation}
 D_x^m\cdots D_y^n f\cdot g = \left.
(\pa_x-\pa_{x'})^m\cdots(\pa_y-\pa_{y'})^n
f(x,\dots,y)g(x',\dots,y')\right|_{x'=x,y'=y}, 
\end{equation}
and the bar $\bar{\:\cdot\:}$ denotes complex conjugation. 
Based on the bilinear equations \eqref{Bi2dNLS}, 
Sasa, Ohta and Matsukidaira \cite{SOM} constructed determinant-type 
solutions. Their work strongly suggests that equation \eqref{2dNLS} 
may be related to the KP hierarchy. 

The main purpose of the present paper is to generalize the results of 
the previous work \cite{IT} so that we can treat equation
\eqref{2dNLS} and the SDYM equation. 
We shall use the language of formal pseudo-differential 
operators (PsDO for short) that have matrix coefficients. 
In other words, we will generalize the theory of the multi-component 
KP hierarchy \cite{Dickey,Sato,UT} to the $(2+1)$-dimensional NLS hierarchy. 
We will also use the free fermion operators \cite{DJKM,JM,JMD} 
to clarify the relation to the toroidal Lie algebras. 

This paper is organized as follows: In Section 2, we introduce SDYM-type time
evolutions to the $2$-component KP hierarchy and show that the resulting
hierarchy contains the $(2+1)$-dimensional NLS equation \eqref{2dNLS}. 
We also discuss bilinear identity for the $\tau$-functions, and relation to
the SDYM equation. 
In Section 3, we present two ways to construct special solutions. 
Relation to toroidal Lie algebras is explained in Section 4. 
Based on the Fock space representation, we derive the bilinear 
identities from the representation-theoretical viewpoint. 
Section 5 is devoted to the concluding remarks. 

\section{Formulation of the $(2+1)$-dimensional NLS hierarchy}
\setcounter{equation}{0}%
\subsection{$2$-component KP hierarchy}
We first review the theory of the multi-component KP hierarchy 
\cite{Dickey,Sato,UT} in the language of formal pseudo-differential
operators with $(N\times N)$-matrix coefficients. 

Let $\pa_x$ denote the derivation $\pa/\pa x$.  
A formal PsDO is a formal linear combination, 
$\hat{\bm{A}}=\sum_n \bm{a}_n \pa_x^n$, of integer powers of $\pa_x$ 
with matrix coefficients $\bm{a}_n = \bm{a}_n(x)$ that depend on $x$.  
The index $n$ ranges over all integers with an upper 
bound.  The least upper bound is called the 
{\it order} of this PsDO.  The first non-vanishing 
coefficient $\bm{a}_N$ is called the {\it leading 
coefficient}.  If the leading coefficient is equal 
to $\bm{I}$, the unit matrix, the PsDO is said to be {\it monic}.  
It is convenient to use the following notation: 
\begin{equation}
 [\hat{\bm{A}}]_{\ge 0} \defeq \sum_{n \ge 0} \bm{a}_n \pa_x^n, \quad
 [\hat{\bm{A}}]_{< 0} \defeq \sum_{n<0} \bm{a}_n \pa_x^n, \quad
 (\hat{\bm{A}})_k \defeq \bm{a}_k. 
\end{equation}

Addition and multiplication (or composition) of 
two PsDO's are defined as follows.  Addition of 
two PsDO's is an obvious operation, namely, the 
termwise sum of the coefficients.  Multiplication 
is defined by extrapolating the Leibniz rule 
\begin{equation}
\pa_x^n \circ f = \sum_{k \ge 0} \binom{n}{k} f^{(k)} \pa_x^{n - k},
\end{equation}
to the case where $n$ is negative.  Here the circle ``$\circ$'' 
stands for composition of two operators, and 
$f^{(k)}$ the $k$-th derivative $\pa^k f/\pa x^k$ 
of $f$.  More explicitly, the product
$\hat{\bm{C}}=\hat{\bm{A}}\circ\hat{\bm{B}}$ of two PsDO's
$\hat{\bm{A}}=\sum_n\bm{a}_n\pa_x^n$ and 
$\hat{\bm{B}}=\sum_n\bm{b}_n\pa_x^n$ is given by 
\begin{equation}
\hat{\bm{C}} = \sum_{m,n,k} \binom{m}{k} 
\bm{a}_m^{(k)} \bm{b}_n \pa_x^{m+n-k}
\end{equation}
Note that the $n$-th order coefficient $\bm{c}_n = (\hat{\bm{C}})_n$ 
is the sum of a finite number of terms.  
Any PsDO $\hat{\bm{A}}=\sum_{n \le N}\bm{a}_n\pa_x^n$ with an invertible
leading coefficient $\bm{a}_N$ has an inverse PsDO. In particular, any
monic PsDO is invertible.  We shall frequently write
$\hat{\bm{A}}\hat{\bm{B}}$ rather than $\hat{\bm{A}}\circ\hat{\bm{B}}$ 
if it does not cause confusion. 
One can make sense of the action of PsDO's on $\ee^{\lambda x}$ by 
simply extrapolating the derivation rule 
$
 \pa_x^n \ee^{\lambda x} = \lambda^n \ee^{\lambda x}
$
to negative powers of $\pa_x$. 

Hereafter we consider only the $2$-component case since it is sufficient
for our purpose. However it is easy to generalize the results below to
higher-component case. 
Let us introduce the $2$-component version of the {\it Sato-Wilson
operator}, 
\begin{equation}
\hat{\bm{W}} \defeq \bm{I} + \sum_{n=1}^\infty \bm{w}_n \pa_x^{-n}, 
\end{equation}
where $\bm{w}_j=\bm{w}_j(x,\ul{x}^{(1)},\ul{x}^{(2)})$ denote the
$(2\times 2)$-matrix-valued functions that depend on infinitely many
variables $(x,\ul{x}^{(1)},\ul{x}^{(2)})=
(x,x_1^{(1)},x_2^{(1)},\ldots,x_1^{(2)},x_2^{(2)},\ldots)$. 
The $2$-component KP hierarchy is defined by the {\it Sato equation}, 
\begin{equation}
\frac{\pa\hat{\bm{W}}}{\pa x^{(\alpha)}_n} = 
\hat{\bm{B}}^{(\alpha)}_n\hat{\bm{W}}-\hat{\bm{W}}\bm{E}_{\alpha} \pa_x^n, 
\qquad
\hat{\bm{B}}^{(\alpha)}_n = \left[ 
\hat{\bm{W}}\bm{E}_{\alpha}\pa_x^n\hat{\bm{W}}^{-1} \right]_{\geq 0},
\label{2cSatox}
\end{equation}
for $n=1,2,\dots$, $\alpha = 1,2$, 
with $\bm{E}_{\alpha}=(\delta_{i\alpha}\delta_{j\alpha})_{i,j=1,2}$. 

\subsection{From the $2$-component KP hierarchy to the $(2+1)$-dimensional
NLS hierarchy} 
We impose the constraint $[\hat{\bm{W}}\pa_x\hat{\bm{W}}^{-1}]_{<0}=0$, 
which means that 
\begin{equation}
\label{red1}
\hat{\bm{W}}\pa_x\hat{\bm{W}}^{-1} = \bm{I}\pa_x, 
\end{equation}
or equivalently, 
\begin{equation}
\label{reduction}
\frac{\pa\hat{\bm{W}}}{\pa x}=0, \qquad
\left(\frac{\pa}{\pa x^{(1)}_n}+\frac{\pa}{\pa x^{(2)}_n}\right)
\hat{\bm{W}}=0.
\end{equation}
If a PsDO $\hat{\bm{A}}$ satisfies the condition 
$[\pa_x, \hat{\bm{A}}]=0$, the correspondence 
\begin{equation}
\label{corresp}
\hat{\bm{A}}=\sum_{n\in\IZ}\bm{a}_n\pa_x^n \quad\leftrightarrow\quad 
\bm{A}(\lambda)=\sum_{n\in\IZ}\bm{a}_n\lambda^n
\end{equation}
preserves sums, products and commutators. Here $\lambda$ is used as 
a formal indeterminate ({\it spectral parameter}). 
Under this constraint, we can use the correspondence \eqref{corresp} 
and the remaining time evolutions are those of 
$x_n\defeq x^{(1)}_n-x^{(2)}_n$. The evolution equations with respect 
to $x_n$ are obtained from \eqref{2cSatox}: 
\begin{equation}
\label{RedSato}
\begin{aligned}
\frac{\pa \bm{W}(\lambda)}{\pa x_n} 
& = \bm{B}_n(\lambda)\bm{W}(\lambda) - \lambda^n \bm{W}(\lambda)\bm{Q}, \\
\bm{B}_n(\lambda) &\defeq \bm{B}^{(1)}_n(\lambda)-\bm{B}^{(2)}_n(\lambda)
=\left[ \lambda^n\bm{W}(\lambda)\bm{Q}\bm{W}(\lambda)^{-1} 
\right]_{\geq 0},
\end{aligned}
\end{equation}
with $\bm{Q}=\bm{E}_1-\bm{E}_2$. 
For example, the explicit form of 
$\bm{B}_1(\lambda)$ and $\bm{B}_2(\lambda)$ are given by 
\begin{equation}
\begin{aligned}
\bm{B}_1(\lambda) &= \lambda\bm{Q} + \bm{w}_1\bm{Q}-\bm{Q}\bm{w}_1,\\
\bm{B}_2(\lambda) &= \lambda^2 \bm{Q}
 + \lambda (\bm{w}_1\bm{Q}-\bm{Q}\bm{w}_1)
 +(\bm{w}_2\bm{Q}-\bm{Q}\bm{w}_2)
 -(\bm{w}_1\bm{Q}-\bm{Q}\bm{w}_1)\bm{w}_1.
\end{aligned}
\end{equation}

We now introduce a new set of infinite time variables 
$\ul{y}=(y_0,y_1,\ldots)$. Since the first one $y_0$
plays a special role, we will use the notation 
$\check{y}=(y_1,y_2\ldots)$. 
The time evolutions with respect to $\check{y}$ are defined as 
\begin{equation}
\begin{aligned}
 \frac{\pa \bm{W}(\lambda)}{\pa y_n}
 &= \bm{C}_n(\lambda)\bm{W}(\lambda)
 + \lambda^n \frac{\pa\bm{W}(\lambda)}{\pa y_0}, \\
 \bm{C}_n(\lambda) &= -\left[ \lambda^n 
 \frac{\pa \bm{W}(\lambda)}{\pa y_0}
 \bm{W}(\lambda)^{-1}\right]_{\geq 0}. 
\end{aligned}
\label{2cSatoy}
\end{equation}
We remark that the variables $\ul{y}$ are essentially the same as those 
appeared in the works on hierarchy structure of the SDYM equation 
\cite{N,Tak1,Tak2,Tak3,Tak4}. 

Define a formal series $\bm{\Psi}(\lambda)$ (called the {\it formal 
Baker-Akhiezer function}) as 
\begin{align}
 \label{BAfunc}
 \bm{\Psi}(\lambda) &\defeq \hat{\bm{W}}\bm{\Psi}_0(\lambda)
  = \left(\bm{I} + \sum_{n=1}^\infty \bm{w}_n
 \lambda^{-n}\right)\bm{\Psi}_0(\lambda), \\
\bm{\Psi}_0(\lambda) &\defeq 
 \begin{pmatrix}
  \ee^{\xi(\ul{x};\lambda)+\mu y_0+\mu\xi(\check{y};\lambda)} & 0\\
  0 & \ee^{-\xi(\ul{x};\lambda)+\nu y_0+\nu \xi(\check{y};\lambda)}
 \end{pmatrix}, 
\end{align}
where $\xi(\ul{x};\lambda)$ is given by 
\begin{equation}
 \xi(\ul{x};\lambda) \defeq \sum_{n=1}^\infty x_n \lambda^n. 
\end{equation}
Unlike the ordinary NLS case, the spectral parameter $\lambda=\lambda(\ul{y})$
may depend on the variables $\ul{y}$ as 
\begin{equation}
 \frac{\pa\lambda}{\pa y_n} = \lambda^n\frac{\pa\lambda}{\pa y_0} .
\end{equation}
Both of the additional spectral parameters $\mu$, $\nu$ are constants
with respect to $\ul{x}$ and $\ul{y}$. 
Note that $\bm{\Psi}_0(\lambda)$ obeys linear differential equations, 
\begin{equation}
\frac{\pa}{\pa x_n}\bm{\Psi}_0(\lambda)=\lambda^n\bm{Q}\bm{\Psi}_0(\lambda),
\qquad \frac{\pa}{\pa y_n}\bm{\Psi}_0(\lambda)=
\lambda^n\frac{\pa}{\pa y_0}\bm{\Psi}_0(\lambda). 
\end{equation}

In terms of $\bm{\Psi}(\lambda)$, the evolutions equations
\eqref{RedSato} and \eqref{2cSatoy} are rewritten as 
\begin{align}
 \label{linear_xn}
 \frac{\pa}{\pa x_n}\bm{\Psi}(\lambda) &=
 \bm{B}_n(\lambda)\bm{\Psi}(\lambda), \\
 \label{linear_yn}
 \frac{\pa}{\pa y_n}\bm{\Psi}(\lambda) &= 
 \left( \bm{C}_n(\lambda) + \lambda^n\frac{\pa}{\pa y_0} \right) 
 \bm{\Psi}(\lambda) .
\end{align}
The compatibility condition for \eqref{linear_xn} with $n=1,2$ gives
\begin{equation}
\label{NLS0}
\begin{aligned}
2\frac{\pa w_1^{(12)}}{\pa x_2}
&= \frac{\pa^2 w_1^{(12)}}{\pa x_1^2} +8 (w_1^{(12)})^2 w_1^{(21)}, \\
-2\frac{\pa w_1^{(21)}}{\pa x_2}
&= \frac{\pa^2 w_1^{(21)}}{\pa x_1^2} +8 (w_1^{(21)})^2 w_1^{(12)}, 
\end{aligned}
\end{equation}
where $w_1^{(ij)}$ denotes the $(i,j)$-element of the matrix $\bm{w}_1$. 
If we set $u=-2w_1^{(12)}$, $\overline{u}=2w_1^{(21)}$, 
$t_1=\ii x_1$  and $t_2=2\ii x_2$, then equations \eqref{NLS0} are
reduced to the NLS equation \eqref{NLS}. 

In the case of $n=1$, equations \eqref{linear_xn}, \eqref{linear_yn} can
be written explicitly as
\begin{align}
\label{linear_x1}
\frac{\pa}{\pa x_1}\bm{\Psi}(\lambda) &= 
( \lambda \bm{Q} + \bm{w}_1\bm{Q}-\bm{Q}\bm{w}_1)\bm{\Psi}(\lambda), \\
\label{linear_y1}
\frac{\pa}{\pa y_1}\bm{\Psi}(\lambda) &= 
\left( \lambda\frac{\pa}{\pa y_0}-\frac{\pa \bm{w}_1}{\pa y_0}\right)
\bm{\Psi}(\lambda) .
\end{align}
The compatibility condition for \eqref{linear_x1} and \eqref{linear_y1}
is reduced to the following nonlinear coupled equations: 
\begin{equation}
\label{2dNLS0}
\begin{aligned}
- \pa_{y_1} & w_1^{(12)} = 
-\pa_{x_1}\pa_{y_0} w_1^{(12)} + w_1^{(12)}\cdot 
\pa_{y_0}(w_1^{(11)}-w_1^{(22)}) , \\
\pa_{y_1} & w_1^{(21)} = 
-\pa_{x_1}\pa_{y_0} w_1^{(21)} + w_1^{(21)}\cdot 
\pa_{y_0}(w_1^{(11)}-w_1^{(22)}) , \\
\pa_{x_1} & (w_1^{(11)}-w_1^{(22)}) = 
-2 (w_1^{(12)}w_1^{(21)}) .
\end{aligned}
\end{equation}
If we impose the conditions
\begin{equation}
\label{reality0}
w_1^{(21)} = -\overline{w_1^{(12)}}, \quad 
w_1^{(22)} = \overline{w_1^{(11)}}, \quad 
x_j\in\ii\IR, \quad y_j\in\IR, 
\end{equation}
equations \eqref{2dNLS0} yield the $(2+1)$-dimensional NLS
equation \eqref{2dNLS} by setting $u=w_1^{(12)}$, $x=\ii x_1$, $y=y_0$,
$t=-y_1$. In this sense, the evolution equations \eqref{2cSatox} and
\eqref{2cSatoy}, with the reduction condition \eqref{red1}, give a 
hierarchy of integrable equations associated with the $(2+1)$-dimensional NLS
equation \eqref{2dNLS}. 
We note that the linear equations \eqref{linear_x1}, \eqref{linear_y1}
are the Lax pair that has been given in several preceding works 
\cite{Bog2,Sc,St1,St2}. 

\subsection{Relation to the self-dual Yang-Mills equation}
We first briefly review the classical theory of the self-dual gauge fields
\cite{Pra,Yang}. 
Let $\bm{A}_u=\bm{A}_u(y,z,\by,\bz)$ ($u=y,z,\by,\bz$) be matrix-valued
functions. Here the suffix does {\it not} denotes the differentiation. 
Define the {\it field strength} $\bm{F}_{uv}$ ($u,v=y,z,\by,\bz$) as 
\begin{equation}
\bm{F}_{uv}=\pa_u \bm{A}_v -\pa_v \bm{A}_u + [\bm{A}_u, \bm{A}_v]. 
\end{equation}
The self-dual Yang-Mills (SDYM) equations is formulated as 
\begin{equation}
\label{SDYM1}
\bm{F}_{yz}=\bm{F}_{\by\bz}=0, \qquad \bm{F}_{y\by}+\bm{F}_{z\bz}=0, 
\end{equation}
which is invariant under the {\it gauge-transformation}
\begin{equation}
\label{GaugeTr}
\bm{A}_u \mapsto \tilde{\bm{A}}_u
 = \bm{G}^{-1}\bm{A}_u\bm{G} + \bm{G}^{-1}(\pa_u \bm{G}). 
\end{equation}

Under the suitable choice of $\bm{G}$ of \eqref{GaugeTr}, we can take 
$\bm{A}_y=\bm{A}_z=0$ and the self-duality equations \eqref{SDYM1} is
reduced to 
\begin{equation}
\label{SDYM3}
\pa_{\by} \bm{A}_{\bz} -\pa_{\bz} \bm{A}_{\by}
 + [\bm{A}_{\by}, \bm{A}_{\bz}]=0, \qquad
\pa_y \bm{A}_{\by} +\pa_z \bm{A}_{\bz} =0. 
\end{equation}
The second equation ensures the existence of the potential $w$ 
such that 
\begin{equation}
\bm{A}_{\by} = - \pa_z \bm{w}, \qquad \bm{A}_{\bz}=\pa_y \bm{w}. 
\end{equation}
In terms of $\bm{w}$, we can rewrite \eqref{SDYM3} as
\begin{equation}
\label{SDYMw}
(\pa_y\pa_{\by}+\pa_z\pa_{\bz})\bm{w} + [\pa_y \bm{w}, \pa_z \bm{w}] =0. 
\end{equation}
We note that this equation appeared in several works on the SDYM 
\cite{BLR,LM,Parkes} and is associated with a cubic action \cite{LM,Parkes}. 
The nonlinear equations \eqref{SDYMw} can be obtained as the
compatibility condition for the following linear equations: 
\begin{equation}
\label{linSDYM}
(\pa_{\bz}-\lambda\pa_y + \pa_y \bm{w})\bm{\Psi}=0, \qquad
(\pa_{\by}+\lambda\pa_z - \pa_z \bm{w})\bm{\Psi}=0. 
\end{equation}
These equations are of the same form as \eqref{linear_y1}. 
To treat these equations simultaneously, we introduce another set of
variables $\ul{z}=(z_0,z_1,z_2,\ldots)$, which play the same role as
$\ul{y}$, i.e., 
\begin{equation}
\label{linear_zn}
\frac{\pa}{\pa z_n}\bm{\Psi}(\lambda) = 
\left( \hat{\bm{C}}_n(\lambda) + \lambda^n\frac{\pa}{\pa z_0} \right) 
\bm{\Psi}(\lambda) .
\end{equation}
In particular, the evolution equation with respect to $z_1$ is
\begin{equation}
\label{linear_z1}
\frac{\pa}{\pa z_1}\bm{\Psi}(\lambda) = 
\left( \lambda\frac{\pa}{\pa z_0}-\frac{\pa \bm{w}_1}{\pa z_0}\right)
\bm{\Psi}(\lambda) .
\end{equation}
Setting $y_0=y$, $y_1=\bz$, $z_0=z$, $z_1=-\by$, and $\bm{w}_1=\bm{w}$, 
we can identify \eqref{linear_y1} and \eqref{linear_z1} with 
the linear equations \eqref{linSDYM} for the SDYM. 

\subsection{Bilinear identity}
\begin{theorem}
\label{BIforBA}
The formal Baker-Akhiezer functions $\bm{\Psi}(\lambda;\ul{x},\ul{y})$ 
satisfy the bilinear equation, 
\begin{equation}
\label{BI0}
\oint\frac{\dd\lambda}{2\pi\ii} \, \lambda^k 
\bm{\Psi}(\lambda;\ul{x},y_0-\xi(\check{b},\lambda),\check{y}+\check{b})
\bm{\Psi}(\lambda;\ul{x}',y_0-\xi(\check{c},\lambda),\check{y}+\check{c})^{-1}
= \bm{0}, 
\end{equation}
for $k\ge 0$. Here $\ul{x}$, $\ul{x}'$, $\ul{y}$, $\check{b}$ and 
$\check{c}$ are understood to be independent variables. 
The contour integral is understood symbolically, 
namely, just to extract the coefficient of $\lambda^{-1}: 
\oint \lambda^n \dd\lambda/(2\pi\ii) = \delta_{n,-1}$. 
\end{theorem}
\begin{proof}
In the case of $\ul{x}'=\ul{x}$, $\check{b}=\check{c}=0$, 
it is clear that  
\begin{equation}
\label{trivial}
\oint\frac{\dd\lambda}{2\pi\ii} \, \lambda^k 
\bm{\Psi}(\lambda;\ul{x},\ul{y})
\bm{\Psi}(\lambda;\ul{x},\ul{y})^{-1}
= \bm{0}, 
\end{equation}
for $k\ge 0$. 
Iteration of the evolution equation of 
$\bm{\Psi}(\lambda;\ul{x},\ul{y})$ gives rise to higher order equations
of the form, 
\begin{equation}
\pa_{x_1}^{\alpha_1}\pa_{x_2}^{\alpha_2}\dots
\bm{\Psi}(\lambda;\ul{x},\ul{y})
= \bm{B}_{\alpha_1,\alpha_2,\ldots}(\lambda)\bm{\Psi}(\lambda;\ul{x},\ul{y}), 
\end{equation}
for $k,\alpha_1,\alpha_2,\ldots \ge 0$, and 
$\bm{B}_{\alpha_1,\alpha_2,\ldots}(\lambda)$ 
being a polynomial in $\lambda$. 
Combining these equations with \eqref{trivial}, 
we obtain the bilinear equations, 
\begin{equation}
\oint\frac{\dd\lambda}{2\pi\ii} \: \lambda^k
\left(\pa_{x_1}^{\alpha_1}\pa_{x_2}^{\alpha_2}\dots
\bm{\Psi}(\lambda;\ul{x},\ul{y})\right)
\bm{\Psi}(\lambda;\ul{x},\ul{y})^{-1} = \bm{0}, 
\end{equation}
which can be cast into a single equation, 
\begin{equation}
\oint\frac{\dd\lambda}{2\pi\ii}\: \lambda^k
\bm{\Psi}(\lambda;\ul{x},\ul{y})
\bm{\Psi}(\lambda;\ul{x}',\ul{y})^{-1} = \bm{0}.
\end{equation}

Next we use \eqref{linear_yn} to obtain
\begin{equation}
(\pa_{y_1} - \lambda\pa_{y_0})^{\beta_1} 
(\pa_{y_2} - \lambda^2\pa_{y_0})^{\beta_2}\dots
\bm{\Psi}(\lambda;\ul{x},\ul{y}) 
= \bm{C}_{\beta_1,\beta_2,\ldots}(\lambda)
\bm{\Psi}(\lambda;\ul{x},\ul{y}) 
\end{equation}
for $\beta_1,\beta_2,\ldots \ge 0$, 
$\bm{C}_{\beta_1,\beta_2,\ldots}(\lambda)$ 
being a polynomial in $\lambda$. This yields 
\begin{equation}
\oint\frac{\dd\lambda}{2\pi\ii} \: \lambda^k
\left( (\pa_{y_1} - \lambda\pa_{y_0})^{\beta_1} 
(\pa_{y_2} - \lambda^2\pa_{y_0})^{\beta_2}\dots
\bm{\Psi}(\lambda;\ul{x},\ul{y})\right)
\bm{\Psi}(\lambda;\ul{x}',\ul{y})^{-1} = \bm{0},
\end{equation}
and we have 
\begin{equation}
\oint\frac{\dd\lambda}{2\pi\ii}\: \lambda^k
\bm{\Psi}(\lambda;\ul{x},y_0-\xi(\check{b},\lambda),\check{y}+\check{b})
\bm{\Psi}(\lambda;\ul{x}',y_0,\check{y})^{-1} = \bm{0}.
\end{equation}
Similar discussion with the differential equations for 
$\bm{\Psi}(\lambda)^{-1}$ gives the desirous result.
\end{proof}

Now we derive the bilinear identity for $\tau$-functions of the
$(2+1)$-dimensional NLS hierarchy. 
In the $2$-component case \cite{Dickey,JM,UT}, we need three
$\tau$-functions $F(\ul{x},\ul{y})$, $G(\ul{x},\ul{y})$ and
$\tilde{G}(\ul{x},\ul{y})$ that are consistently introduced by 
\begin{align}
\bm{\Psi}(\lambda) &= \frac{1}{F(\ul{x},\ul{y})}
\label{defFFGtil}\\
& \times\begin{pmatrix}
F(x^{(1)}-[\lambda^{-1}],x^{(2)},\ul{y}) & 
\lambda^{-1}G(x^{(1)},x^{(2)}-[\lambda^{-1}],\ul{y}) \\
\lambda^{-1}\tilde{G}(x^{(1)}-[\lambda^{-1}],x^{(2)},\ul{y}) & 
F(x^{(1)},x^{(2)}-[\lambda^{-1}],\ul{y}) 
\end{pmatrix}
\bm{\Psi}_0(\lambda), \nonumber\\
\bm{\Psi}(\lambda)^{-1} &= 
\frac{1}{F(\ul{x},\ul{y})}\bm{\Psi}_0(\lambda)^{-1}\\
& \times\begin{pmatrix}
F(x^{(1)}+[\lambda^{-1}],x^{(2)},\ul{y}) & 
-\lambda^{-1}G(x^{(1)}+[\lambda^{-1}],x^{(2)},\ul{y}) \\
-\lambda^{-1}\tilde{G}(x^{(1)},x^{(2)}+[\lambda^{-1}],\ul{y}) &
F(x^{(1)},x^{(2)}+[\lambda^{-1}],\ul{y}) 
\end{pmatrix}
 , \nonumber
\end{align}
where we have used the notation 
$[\lambda^{-1}] \defeq \left(\,
1/\lambda,1/2\lambda^2,1/3\lambda^3,\ldots \,\right)$. 
For the moment, we will forget the complex structure, i.e., $\tilde{G}$
is not assumed to be the complex conjugate of $G$. 
Note that $F$, $G$ and $\tilde{G}$ depend only on
$x_n=x_n^{(1)}-x_n^{(2)}$ under the condition \eqref{red1}. 
The denominator of the integral of \eqref{BI0} is of the form 
\begin{equation}
F(\ul{x},y_0-\xi(\check{b},\lambda),\check{y}+\check{b})
F(\ul{x}',y_0-\xi(\check{c},\lambda),\check{y}+\check{c}), 
\label{FF}
\end{equation}
which is a power series.  According to Theorem \ref{BIforBA}, 
one can insert any power series of $\lambda$ in 
\eqref{BI0}.  If we insert \eqref{FF} itself therein, the 
denominator cancels out, so that we obtain the 
following identities for the $(2+1)$-dimensional 
NLS hierarchy:
\begin{corollary}
\label{Cor:BI_2+1NLS}
For any non-negative integer $k$, the functions $F$, $G$ and $\tilde{G}$ 
satisfy the bilinear equations, 
\begin{align}
\oint & \frac{\dd\lambda}{2\pi\ii} \: \Big\{ 
\lambda^k\ee^{\xi((\ul{x}-\ul{x}')/2,\lambda)} 
F(\ul{x}-[\lambda^{-1}],\ul{y}+\ul{b}_{\lambda})
F(\ul{x}'+[\lambda^{-1}],\ul{y}+\ul{c}_{\lambda}) \nonumber\\
& \quad - \lambda^{k-2} \ee^{\xi((\ul{x}'-\ul{x})/2,\lambda)} 
G(\ul{x}+[\lambda^{-1}],\ul{y}+\ul{b}_{\lambda})
\tilde{G}(\ul{x}'-[\lambda^{-1}],\ul{y}+\ul{c}_{\lambda})
\Big\} = 0, \label{BI_FFGGtil}\\
\oint & \frac{\dd\lambda}{2\pi\ii} \: \lambda^{k-1}\Big\{ 
\ee^{\xi((\ul{x}-\ul{x}')/2,\lambda)} 
F(\ul{x}-[\lambda^{-1}],\ul{y}+\ul{b}_{\lambda})
G(\ul{x}'+[\lambda^{-1}],\ul{y}+\ul{c}_{\lambda}) \nonumber\\
& \qquad\quad - \ee^{\xi((\ul{x}'-\ul{x})/2,\lambda)} 
G(\ul{x}+[\lambda^{-1}],\ul{y}+\ul{b}_{\lambda})
F(\ul{x}'-[\lambda^{-1}],\ul{y}+\ul{c}_{\lambda})
\Big\} = 0, \label{BI_GF}\\
\oint & \frac{\dd\lambda}{2\pi\ii} \: \lambda^{k-1}\Big\{ 
\ee^{\xi((\ul{x}-\ul{x}')/2,\lambda)} 
\tilde{G}(\ul{x}-[\lambda^{-1}],\ul{y}+\ul{b}_{\lambda})
F(\ul{x}'+[\lambda^{-1}],\ul{y}+\ul{c}_{\lambda}) \nonumber\\
& \qquad\quad - \ee^{\xi((\ul{x}'-\ul{x})/2,\lambda)} 
F(\ul{x}+[\lambda^{-1}],\ul{y}+\ul{b}_{\lambda})
\tilde{G}(\ul{x}'-[\lambda^{-1}],\ul{y}+\ul{c}_{\lambda})
\Big\} = 0, \label{BI_GtilF}
\end{align}
where $\ul{b}_{\lambda}$ denotes $(b_0,b_1,b_2,\ldots)$ with the
constraint $b_0=-\xi(\check{b},\lambda)$. 
\end{corollary}

The bilinear identities \eqref{BI_FFGGtil}--\eqref{BI_GtilF} 
can be rewritten into a series of Hirota-type
differential equations. The simplest examples are
\begin{eqnarray}
D_{x_1}^2 F\cdot F - 2G\tilde{G} & =0, \label{bilNLS1}\\
(D_{x_1}D_{y_0}-D_{y_1})G\cdot F & =0, \label{bilNLS2}\\
(D_{x_1}D_{y_0}+D_{y_1})\tilde{G}\cdot F & =0. \label{bilNLS3}
\end{eqnarray}
These equations coincide with \eqref{Bi2dNLS} if we set
$\tilde{G}=-\bar{G}$, $x_1=-\ii X$, $y_0=Y$, and $y_1=T$. 

\section{Special solutions of the hierarchy}
\setcounter{equation}{0}%
\subsection{Double-Wronskian solutions}
We first apply the method due to one of the authors 
\cite{Tak1,Tak2,Tak3} to construct
a special class of solutions for the $(2+1)$-dimensional NLS hierarchy, 
which we shall seek in the form 
\begin{equation}
\bm{\Psi}(\lambda)
=\left(
\bm{I}\lambda^N +\bm{w}_1 \lambda^{N-1}+\cdots+\bm{w}_N
\right)\bm{\Psi}_0, 
\end{equation}
with $\bm{w}_n=\bm{w}_n(\ul{x},\ul{y})$ being unknown functions.

As the data for the solution constructed below, 
let us consider a formal series 
$\bm{\Xi}(\lambda)=\sum_{j\in\IZ}\bm{\xi}_j\lambda^{-j}$ 
where $\bm{\xi}_j=\bm{\xi}_j(\ul{x},\ul{y})$ are 
$(2\times 2N)$-matrix-valued functions of the form, 
\begin{equation}
\bm{\xi}_j(\ul{x},\ul{y})=
\begin{pmatrix}
f^{(j)}_1(\ul{x},\ul{y}) & \cdots & 
f^{(j)}_{2N}(\ul{x},\ul{y})\\
g^{(j)}_1(\ul{x},\ul{y}) & \cdots & 
g^{(j)}_{2N}(\ul{x},\ul{y})
\end{pmatrix}. \label{bmf}
\end{equation}
Here we assume 
\begin{equation}
\label{independence}
\det\begin{pmatrix}
f^{(0)}_1 & \cdots & f^{(N)}_1 & g^{(0)}_1 & \cdots & g^{(N)}_1\\
   \vdots & \ddots &    \vdots &    \vdots & \ddots & \vdots \\
f^{(0)}_{2N} & \cdots & f^{(N)}_{2N} & g^{(0)}_{2N} & \cdots & g^{(N)}_{2N}
\end{pmatrix} \neq 0 .
\end{equation}
We furthermore impose the following conditions for
$\bm{\Xi}(\lambda)$: 
\begin{eqnarray}
& \bullet & \frac{\pa}{\pa x_n}\bm{\Xi}(\lambda) = 
\lambda^n\bm{Q}\bm{\Xi}(\lambda) + \bm{\Xi}(\lambda)\bm{\alpha}_n
\quad (n=1,2,\ldots), \label{x}\\
& \bullet & \frac{\pa}{\pa y_n}\bm{\Xi}(\lambda) = 
\lambda^n\frac{\pa}{\pa y_0}\bm{\Xi}(\lambda)
 + \bm{\Xi}(\lambda)\bm{\beta}_n
\quad (n=1,2,\ldots), \label{y}\\
& \bullet & \lambda \bm{\Xi}(\lambda) = \bm{\Xi}(\lambda)\bm{\gamma}, 
\label{redWN}
\end{eqnarray}
where $\bm{\alpha}_n$, $\bm{\beta}_n$, $\bm{\gamma}$ are 
$(2N\times 2N)$-matrices. 

We now consider a monic polynomial $\bm{W}_N(\lambda)$ of the form 
\begin{equation}
\bm{W}_N(\lambda) = \bm{I}\lambda^N + \bm{w}_1 \lambda^{N-1} + \cdots + \bm{w}_N, 
\end{equation}
which is characterized uniquely by the linear equation 
\begin{equation}
\label{Wf}
\oint\frac{\dd\lambda}{2\pi\ii\lambda}
\bm{W}_N(\lambda) \bm{\Xi}(\lambda) = \bm{0} .
\end{equation}
Solving equation \eqref{Wf} explicitly by the 
Cram\'er formula, we have for example 
\begin{equation}
\label{SolitonSol}
\begin{aligned}
w_1^{(12)} &= (-1)^N
\frac{|0,1,\dots,N;0,1,\dots,N-2|}{|0,1,\dots,N-1;0,1,\dots,N-1|},\\
w_1^{(21)} &= (-1)^{N+1}
\frac{|0,1,\dots,N-2;0,1,\dots,N|}{|0,1,\dots,N-1;0,1,\dots,N-1|}, 
\end{aligned}
\end{equation}
where we have used the notation due to Freeman and Nimmo \cite{F}: 
\begin{equation}
|k_1,\dots,k_m;l_1,\dots,l_n| \defeq
\begin{vmatrix}
f_1^{(k_1)} & \cdots & f_1^{(k_m)} &
g_1^{(l_1)} & \cdots & g_1^{(l_n)} \\
\vdots & \ddots & \vdots & \vdots & \ddots & \vdots \\
f_{2N}^{(k_1)} & \cdots & f_{2N}^{(k_m)} &
g_{2N}^{(l_1)} & \cdots & g_{2N}^{(l_n)} 
\end{vmatrix}. 
\end{equation}

\begin{proposition}
The monic polynomial $\bm{W}_N(\lambda)$ characterized by \eqref{Wf} 
solves \eqref{2cSatox} and \eqref{2cSatoy} simultaneously. 
\end{proposition}
\begin{proof}
{}From \eqref{redWN}, we obtain
\begin{equation}
\label{Wfn}
\oint\frac{\dd\lambda}{2\pi\ii\lambda} \lambda^n 
\bm{W}_N(\lambda) \bm{\Xi}(\lambda) = \bm{0}, 
\end{equation}
for any non-negative integer $n$. 
Differentiating \eqref{Wf} with respect to $x_n$ and applying \eqref{x}, 
we have
\begin{equation}
\oint\frac{\dd\lambda}{2\pi\ii\lambda}
\left( \frac{\pa\bm{W}_N(\lambda)}{\pa x_n}
 + \lambda^n\bm{W}_N(\lambda)\bm{Q} \right) \bm{\Xi}(\lambda) = \bm{0}. 
\label{diffWf}
\end{equation}
There exist polynomials $\bm{B}_n(\lambda)$ and $\bm{R}(\lambda)$ 
such that 
\begin{equation}
\frac{\pa\bm{W}_N(\lambda)}{\pa x_n}
 + \lambda^n\bm{W}_N(\lambda)\bm{Q}
= \bm{B}_n(\lambda) \bm{W}_N(\lambda) + \bm{R}(\lambda),
\label{WN_division}
\end{equation}
where the degree of $\bm{R}(\lambda)$ is at most $N-1$. 
In view of \eqref{Wfn} and \eqref{diffWf}, we obtain 
$\oint\bm{R}(\lambda)\bm{\Xi}(\lambda)\dd\lambda = \bm{0}$. 
The condition \eqref{independence} implies
$\bm{R}(\lambda)=\bm{0}$ and that $\bm{W}_N(\lambda)$ 
satisfies \eqref{RedSato}.

Differentiating \eqref{Wf} with respect to $y_n$ and applying \eqref{y}, 
we have
\begin{equation}
\oint\frac{\dd\lambda}{2\pi\ii\lambda}\left( 
\frac{\pa\bm{W}_N(\lambda)}{\pa y_n} 
   + \lambda^n\bm{W}_N(\lambda)\frac{\pa}{\pa y_0}
\right) \bm{\Xi}(\lambda) = \bm{0}. 
\end{equation}
We can rewrite the second term in the left hand side as follows: 
\begin{equation}
\bm{W}_N(\lambda)\frac{\pa}{\pa y_0}
= \frac{\pa}{\pa y_0}\circ\bm{W}_N(\lambda)
-\frac{\pa \bm{W}_N(\lambda)}{\pa y_0}. 
\end{equation}
Thus we obtain
\begin{equation}
\label{WNSatoy}
\oint\frac{\dd\lambda}{2\pi\ii\lambda}\left( 
\frac{\pa\bm{W}_N(\lambda)}{\pa y_n} 
   - \lambda^n\frac{\pa \bm{W}_N(\lambda)}{\pa y_0}
\right) \bm{\Xi}(\lambda) = \bm{0}. 
\end{equation}
Since the expression in parentheses is an polynomial in $\lambda$, 
we can apply exactly the same argument above to get the unique
polynomial $\bm{C}_n(\lambda)$ and show that 
$\bm{W}_N(\lambda)$ satisfies \eqref{2cSatoy}. 
\end{proof}

Note that equations \eqref{x}--\eqref{redWN} 
are invariant under the transformations 
\begin{eqnarray}
\bm{\Xi}(\lambda) & \mapsto & \bm{\Xi}(\lambda)\bm{H}, \\
\bm{\alpha}_n & \mapsto & 
\bm{H}^{-1}\bm{\alpha}_n\bm{H} + \bm{H}^{-1}\frac{\pa\bm{H}}{\pa x_n}, \\
\bm{\beta}_n & \mapsto & 
\bm{H}^{-1}\bm{\beta}_n\bm{H} + \bm{H}^{-1}\frac{\pa\bm{H}}{\pa y_n}
- \bm{H}^{-1}\bm{\gamma}^n\frac{\pa\bm{H}}{\pa y_0}, \\
\bm{\gamma} & \mapsto & \bm{H}^{-1}\bm{\gamma}\bm{H}, 
\end{eqnarray}
where $\bm{H}=\bm{H}(\ul{x},\ul{y})$ is an $(2N\times 2N)$-invertible
matrix. These formulas are a generalization of the transformations
(2.22) of \cite{Tak2}. As discussed by one of the authors \cite{Tak2}, 
this invariance property shows that the manifold
from which the unknown functions $\{\bm{w}_1,\dots,\bm{w}_N\}$ take
values is essentially a Grassmann manifold. 

We now consider the reality condition \eqref{reality0}: 
\begin{proposition}
\label{Lem:Reality}
Let $\bm{P}_1$ be a $(2\times 2)$-matrix and $\bm{P}_2$ a 
$(2N\times 2N)$-matrix, both of which are invertible. 
If $\bm{\Xi}_j$ satisfies the condition,
\begin{equation}
\label{reality}
\overline{\bm{\Xi}}_j = \bm{P}_1\bm{\Xi}_j\bm{P}_2, 
\end{equation}
then the corresponding  $\bm{W}_N$ satisfies 
\begin{equation}
\bm{P}_1^{-1}\bm{W}_N \bm{P}_1 = \overline{\bm{W}_N}. 
\end{equation}
In particular, if $\bm{P}_1$ is of the form
\begin{equation}
\label{01m10}
\bm{P}=
\begin{pmatrix}
0 & 1\\ -1 & 0
\end{pmatrix}, 
\end{equation}
the coefficients $\bm{w}_j=(w_j^{(ab)})_{a,b=1,2}$ satisfies
\begin{equation}
 w_j^{(22)}=\overline{w_j^{(11)}}, \quad
 w_j^{(21)}=-\overline{w_j^{(12)}} , 
\end{equation}
which agree with \eqref{reality0}. 
\end{proposition}
\begin{proof}
Taking the complex conjugation of \eqref{Wf}, we have 
\begin{equation}
\oint\frac{\dd\lambda}{2\pi\ii\lambda}
(\bm{P}_1^{-1}\overline{\bm{W}_N}(\lambda)\bm{P}_1)
\bm{\Xi}(\lambda) = \bm{0} .
\end{equation}
Then we find that two monic polynomials 
$\bm{W}_N(\lambda)$ and 
$\bm{P}_1^{-1}\overline{\bm{W}_N}(\lambda)\bm{P}_1$ are
characterized by the same data $\bm{\Xi}(\lambda)$. This proves the results.
\end{proof}

We shall give an example that corresponds to soliton-type 
solutions. For the purpose, we choose the form of $f_j$ and $g_j$ of
\eqref{bmf} as 
\begin{equation}
\begin{aligned}
f_k^{(j)}(\ul{x},\ul{y}) &= 
a_k p_k^j\exp\left[ \sum_{n=1}^{\infty}p_k^n x_n
              + r_k y_0 + \sum_{n=1}^{\infty}r_k p_k^n y_n \right], \\
g_k^{(j)}(\ul{x},\ul{y}) &= 
b_k p_k^j\exp\left[ -\sum_{n=1}^{\infty}p_k^n x_n
              + r'_k y_0 + \sum_{n=1}^{\infty}r'_k p_k^n y_n \right], 
\end{aligned}
\end{equation}
where $r_i$, $r'_i$ ($i=1,\dots,N$) are arbitrary complex numbers, and 
$p_i=p_i(\ul{y})$, $a_i=a_i(\ul{y})$, $b_i=b_i(\ul{y})$ ($i=1,\dots,N$) 
are arbitrary (local) solution of the equations
\begin{equation}
\frac{\pa p_i}{\pa y_n} = p_i^n\frac{\pa p_i}{\pa y_0},\quad
\frac{\pa a_i}{\pa y_n} = p_i^n\frac{\pa a_i}{\pa y_0},\quad
\frac{\pa b_i}{\pa y_n} = p_i^n\frac{\pa b_i}{\pa y_0}
\quad (n=1,2,\dots). \label{RiemannWave}
\end{equation}
Moreover, $p_i(\ul{y})$ ($i=1,\dots,N$) are assumed to be 
pairwise distinct. 
Then $\bm{\Xi}(\lambda)$ satisfies the linear equations 
\eqref{x}--\eqref{redWN}. 

Furthermore, if we impose the condition 
\begin{equation}
\begin{aligned}
b_{2j} &= \overline{a_{2j-1}}, \quad a_{2j} = -\overline{b_{2j-1}}, \quad
p_{2j} = \overline{p_{2j-1}}, \\
r_{2j} &= \overline{r'_{2j-1}}, \quad  r'_{2j} = \overline{r_{2j-1}}
\quad (j=1,\dots,N), \\
x_n^{(2)} & =\overline{x_n^{(1)}}, \quad y_n\in\IR 
\quad (n=1,2\ldots), 
\end{aligned}
\end{equation}
then the corresponding $\bm{\Xi}_j$ satisfies \eqref{reality} with
$\bm{P}_1$ of \eqref{01m10}. 
We conclude that the polynomial $\bm{W}_N(\lambda)$ constructed
from the data above gives a solution of the $(2+1)$-dimensional NLS
hierarchy. 
Especially for equations \eqref{2dNLS0}, the solution is given by 
quotient of the ``double Wronskian'' \eqref{SolitonSol}. 

\subsection{Application of the Riemann-Hilbert problem}
In case of the SDYM hierarchy, the Riemann-Hilbert problem plays an
important role \cite{CFYG,Tak4,UN,Ward}. 
We shall show how to apply this problem to the $(2+1)$-dimensional NLS
hierarchy. 

We first consider two solutions $\bm{\Psi}$ and $\bm{\Phi}$ of
\eqref{linear_yn}, which are analytic functions on 
$|\lambda|>1-\epsilon$ (including $\lambda=\infty$) 
and $|\lambda|<1+\epsilon$ respectively. 
Here $\epsilon$ is a constant and $0<\epsilon<1$. 
We further assume that both $\bm{\Psi}$ and $\bm{\Phi}$ are invertible.  
If we define $\bm{g}(\lambda)$ as 
\begin{equation}
\bm{g}(\lambda)=\bm{\Psi}(\lambda)^{-1}\bm{\Phi}(\lambda), 
\end{equation}
then $\bm{g}(\lambda)$ is holomorphic on $1-\epsilon<|\lambda|<1+\epsilon$ and 
satisfies 
\begin{equation}
 \frac{\pa\bm{g}(\lambda)}{\pa y_n} = 
 \lambda^n\frac{\pa\bm{g}(\lambda)}{\pa y_0}. 
\end{equation}
In other words, $\bm{g}=\bm{g}(\lambda;\ul{y})$ is invariant under the
translation, 
\begin{equation}
\label{TransInv}
\bm{g}(\lambda;\ul{y})=\bm{g}(\lambda;\ul{y}+\ul{b}_{\lambda})
\left(=\bm{g}(\lambda;y_0-\xi(\check{b},\lambda),\check{y}+\check{b})\right). 
\end{equation}
On the contrary, starting from $\bm{g}$ with the property
\eqref{TransInv}, we can reconstruct $\bm{\Psi}$ and $\bm{\Phi}$ 
that satisfy the analyticity requirements (the Riemann-Hilbert problem).
Hereafter we assume that $\xi(\check{b},\lambda)$ and $\xi(\check{c},\lambda)$ are
analytic functions on $|\lambda| < 1 + \epsilon$. 
This is a growth condition on the behavior of $b_n$ and $c_n$ as 
$n \to\infty$. 

So far we have not included the NLS-type time evolutions $\ul{x}$. 
To this aim, we define 
$\tilde{\bm{g}}=\tilde{\bm{g}}(\lambda;\ul{x},\ul{y})$ as 
\begin{equation}
\label{defghat}
\tilde{\bm{g}}(\lambda;\ul{x},\ul{y})=\exp[\xi(\ul{x},\lambda)\bm{Q}]
\bm{g}(\lambda;\ul{y})\exp[-\xi(\ul{x},\lambda)\bm{Q}], 
\end{equation}
where we assume $\bm{g}(\lambda;\ul{y})$ enjoys the invariance
\eqref{TransInv}. 
Starting from this $\tilde{\bm{g}}$, we consider the Riemann-Hilbert
decomposition of the matrix $\tilde{\bm{g}}$ such that 
\begin{equation}
\label{RHghat}
\tilde{\bm{g}}(\lambda;\ul{x},\ul{y})=
\tilde{\bm{W}}(\lambda;\ul{x},\ul{y})^{-1}
\tilde{\bm{V}}(\lambda;\ul{x},\ul{y}), 
\end{equation}
where $\tilde{\bm{W}}(\lambda)$ and $\tilde{\bm{V}}(\lambda)$ are analytic
functions on $|\lambda |>1-\epsilon$ and $|\lambda|<1+\epsilon$
respectively. 

\begin{proposition}
If we define $\tilde{\bm{\Psi}}(\lambda)$ as 
\begin{equation}
\tilde{\bm{\Psi}}(\lambda)
 \defeq \tilde{\bm{W}}(\lambda)\exp[\xi(\ul{x},\lambda)\bm{Q}], 
\end{equation}
then $\tilde{\bm{\Psi}}(\lambda)$ solves the bilinear identity 
\eqref{BI0}. 
\end{proposition}
\begin{proof}
The translational invariance 
$\tilde{\bm{g}}(\lambda;\ul{x},\ul{y}+\ul{b}_{\lambda})
=\tilde{\bm{g}}(\lambda;\ul{x},\ul{y}+\ul{c}_{\lambda})$ 
reads 
\begin{equation}
\tilde{\bm{W}}(\lambda;\ul{x},\ul{y}+\ul{b}_{\lambda})
\cdot \tilde{\bm{W}}(\lambda;\ul{x},\ul{y}+\ul{c}_{\lambda})^{-1}
= \tilde{\bm{V}}(\lambda;\ul{x},\ul{y}+\ul{b}_{\lambda})
\cdot \tilde{\bm{V}}(\lambda;\ul{x},\ul{y}+\ul{c}_{\lambda})^{-1}.
\end{equation}
Since the right-hand-side is analytic on $|\lambda|<1+\epsilon$, 
we have 
\begin{equation}
\oint\frac{\dd \lambda}{2\pi\ii}\; \lambda^k
\tilde{\bm{W}}(\lambda;\ul{x},\ul{y}+\ul{b}_{\lambda})\cdot 
\tilde{\bm{W}}(\ul{x},\ul{y}+\ul{c}_{\lambda})^{-1} =0, 
\end{equation}
where the contour is taken as the unit circle with the center at 
$\lambda=0$. 

On the other hand, 
the function $\tilde{\bm{g}}(\lambda;\ul{x},\ul{y})$ satisfies the 
differential equations  
\begin{equation}
\frac{\pa \tilde{\bm{g}}(\lambda;\ul{x},\ul{y})}{\pa x_n}
 = \lambda^n [\bm{Q}, \tilde{\bm{g}}(\lambda;\ul{x},\ul{y})]
\end{equation}
for $n=1,2,\ldots$, which entails 
\begin{equation}
\lambda^n\tilde{\bm{W}}(\lambda)\bm{Q}\tilde{\bm{W}}(\lambda)^{-1}+
\frac{\pa \tilde{\bm{W}}(\lambda)}{\pa x_n}\tilde{\bm{W}}(\lambda)^{-1}=
\lambda^n\tilde{\bm{V}}(\lambda)\bm{Q}\tilde{\bm{V}}(\lambda)^{-1}+
\frac{\pa \tilde{\bm{V}}(\lambda)}{\pa x_n}\tilde{\bm{V}}(\lambda)^{-1}. 
\end{equation}
{}From the analyticity requirements, it follows that the left-hand-side 
is a polynomial of degree at most $n$, which we denote $\bm{B}_n(\lambda)$. 
It is straightforward to shows that 
\begin{equation}
\frac{\pa \tilde{\bm{\Psi}}(\lambda)}{\pa x_n} = 
\bm{B}_n(\lambda)\tilde{\bm{\Psi}}(\lambda), 
\end{equation}
which are nothing but the evolution equations \eqref{linear_xn}. 
With these equations, we can apply the same argument as the proof of 
Theorem \ref{BIforBA}. The resulting equation coincides with \eqref{BI0}.
\end{proof}

Sasa et al. constructed a class of solutions of the $(2+1)$-dimensional
NLS equation \eqref{2dNLS} that are expressed in terms of 
two-directional Wronskians \cite{SOM}. 
In the case of the SDYM, this class of solutions has been discussed 
by Corrigan et al.~\cite{CFYG} based on the Atiyah-Ward ansatz \cite{Ward}, 
\begin{equation}
\bm{g}(\lambda;\ul{y}) = 
\begin{pmatrix}
\lambda^N & \varrho(\lambda;\ul{y})\\
0 & \lambda^{-N}
\end{pmatrix} . 
\end{equation}
Substituting this $\bm{g}$ for \eqref{defghat}, we know that 
$\tilde{\bm{g}}$ is of the same form; 
\begin{equation}
\tilde{\bm{g}}(\lambda;\ul{x},\ul{y}) = 
\begin{pmatrix}
\lambda^N & \tilde{\varrho}(\lambda;\ul{x},\ul{y})\\
0 & \lambda^{-N}
\end{pmatrix}, \quad 
\tilde{\varrho}(\lambda;\ul{x},\ul{y})
 = \varrho(\lambda;\ul{y})\exp[2\xi(\ul{x},\lambda)] . 
\end{equation}
Applying the same argument as that of Corrigan et al.~\cite{CFYG}, 
we can obtain a class of solutions to the $(2+1)$-dimensional NLS
hierarchy, which is an extension of the solutions of Sasa et al.

\section{Relation to the toroidal Lie algebras}
\setcounter{equation}{0}%
\subsection{Definitions and a class of representations}
\label{subsec:DefofTor}
We start with the definitions of the $(M+1)$-toroidal Lie algebra, which is
the universal central extension of the $(M+1)$-fold loop algebra
\cite{Kass,MEY}. 
Let $\frakg$ be a finite-dimensional simple 
Lie algebra over $\IC.$ 
Let $R$ be the ring of Laurent polynomials of 
$M+1$ variables $\IC[s^{\pm 1},t_1^{\pm 1},\dots,t_M^{\pm 1}]$. 
Also assume $M\geq 0$. 
The module of K\"{a}hler differentials 
$\Omega_R$ of $R$ is defined with the canonical 
derivation $\dd:R\rightarrow \Omega_R$.
As an $R$-module, $\Omega_R$ is freely generated by 
$\dd s$, $\dd t_1,\dots,\dd t_M$.
Let $\overline{\cdot}:\Omega_R\rightarrow \Omega_R/\dd R$ 
be the canonical projection.
Let ${\cal K}$ denote $\Omega_R/\dd R.$
Let $(\cdot|\cdot)$ be the normalized Killing form \cite{Kacbook1} on
$\frakg$. 
We define the Lie algebra structure on 
$\gtor\defeq{\frakg}\otimes R\oplus {\cal K}$ by
\begin{equation}
  [X\otimes f,Y\otimes g]=[X,Y]\otimes fg+(X|Y)\overline{(\dd f)g},\qquad
  [{\cal K},\gtor]=0 .
\label{bracket-tor}
\end{equation}
This bracket defines a universal central extension 
of $\frakg\otimes R$ \cite{Kass,MEY}. 

We have, for $u=s,t_1,\dots,t_M$, the Lie subalgebras
\begin{equation}
 \widehat{\frakg}_u\defeq\frakg\otimes{\IC}[u^{\pm 1}]
 \oplus{\IC}\,\overline{\dd\log u}, 
\end{equation}
with the brackets given by
\begin{equation}
\label{affineRel}
 [X\otimes u^m,Y\otimes u^n]
 = [X,Y]\otimes u^{m+n}+m\delta_{m+n,0}(X|Y)\, K_u,
\end{equation}
which are isomorphic to the affine Lie algebra $\widehat{\frakg}$
with the canonical central element $K_u\defeq\overline{\dd\log u}$. 
In terms of the generating series, 
\begin{equation}
 X(z)\defeq \sum_{n\in\IZ}X\otimes u^n \cdot z^{-n-1}, 
\end{equation}
the relation \eqref{affineRel} is equivalent to the following {\it operator
product expansion} (OPE, in short. See, for example, \cite{Kacbook2}) : 
\begin{equation}
\label{affineOPE}
X(z)Y(w) \sim  \frac{1}{z-w}[X,Y](w) + \frac{1}{(z-w)^2}(X|Y)K_u.
\end{equation}

We prepare the generating series of $\gtor$ as follows:
\begin{align}
 A_{\ul{m}}(z) &\defeq 
   \sum_{n\in\IZ}A\otimes s^n\ul{t}^{\ul{m}}\cdot z^{-n-1},\\
 K_{\ul{m}}^s(z) &\defeq 
   \sum_{n\in\IZ}\overline{s^n\ul{t}^{\ul{m}}\,\dd\log s}\cdot z^{-n},\\
 K_{\ul{m}}^{t_k}(z) &\defeq
   \sum_{n\in\IZ}\overline{s^n\ul{t}^{\ul{m}}\,\dd\log t_k}\cdot z^{-n-1}, 
\end{align}
where $A\in\frakg$, $\ul{m}=(m_1,\dots,m_M)\in\IZ^M$, 
$\ul{t}^{\ul{m}}=t_1^{m_1}\cdots t_M^{m_M}$, and $k=1,\dots,M$.
The relation $\overline{\dd (s^n\ul{t}^{\ul{m}})}=0$ can be neatly expressed by 
these generating series as  
\begin{equation}
 \frac{\pa}{\pa z}K_{\ul{m}}^s(z)=\sum_{k=1}^{M}m_k K_{\ul{m}}^{\ul{t}}(z), 
\end{equation}
and the bracket \eqref{bracket-tor} as 
\begin{eqnarray}
X_{\ul{m}}(z)Y_{\ul{n}}(w) &\sim&  
\frac{1}{z-w}[X,Y]_{\ul{m}+\ul{n}}(w) 
 + \frac{1}{(z-w)^2}(X|Y)K_{\ul{m}+\ul{n}}^s(w) \nonumber\\
&& \qquad + \sum_{k=1}^M\frac{m_k}{z-w}(X|Y)K_{\ul{m}+\ul{n}}^{t_k}(w) .
\label{torOPE}
\end{eqnarray}

To construct a class of representations of $\gtor$, we consider the space of
polynomials, 
\begin{equation}
F_y \defeq \mathop{\otimes}_{k=1}^M \left(
\IC[y_j^{(k)},\, j\in\IN]\otimes\IC[\ee^{\pm y^{(k)}_0}]\right). 
\end{equation}
We define the generating series
\begin{equation}
 \varphi^{(k)}(z) \defeq \sum_{n\in\IN}n y_n^{(k)}z^{n-1}, \quad
 V_{\ul{m}}(\ul{y};z) \defeq \prod_{k=1}^M 
 \exp\left[ m_k\sum_{n\in\IN}y_n^{(k)}z^n \right], 
\end{equation}
for each $k=1,\dots,M$, $\ul{m}\in\IZ^M$. 
\begin{proposition}
\label{prop:tor_ext}
Let $(V,\pi)$ be a representation of $ \widehat\frakg_s$
such that $\overline{\dd\log s}\mapsto c\cdot\mathrm{id}_V$ for $c\in\IC$.
Then we can define the representation $\pi^{\mathrm{tor}}$ of $\gtor$ on
$V\otimes F_y$ such that 
\begin{eqnarray}
 X_{\ul{m}}(z) &\mapsto& X^\pi(z)\otimes V_{\ul{m}}(z), \\
 K_{\ul{m}}^s(z) &\mapsto& c\cdot\mathrm{id}_V \otimes V_{\ul{m}}(z), \\
 K_{\ul{m}}^{t_k}(z) &\mapsto&
    c\cdot\mathrm{id}_V \otimes \varphi^{(k)}(z)V_{\ul{m}}(z),
\end{eqnarray}
where $X\in\frakg$, $m\in\IZ$ and
 $X^\pi(z)\defeq\sum_{n\in\IZ}\pi(X\otimes s^n)z^{-n-1}$.
\end{proposition}
\begin{proof}
By the OPE \eqref{affineOPE} and the property 
$V_{\ul{m}}(z)V_{\ul{n}}(z)=V_{\ul{m}+\ul{n}}(z)$, we obtain 
\begin{eqnarray}
\lefteqn{\left(X(z)\otimes 
V_{\ul{m}}(z)\right)\left(Y(w)\otimes V_{\ul{n}}(w)\right)}\qquad
\nonumber\\
 & \sim &
\left\{\frac{1}{z-w}[X,Y](w) + \frac{c}{(z-w)^2}(X|Y)\right\}
\nonumber\\
& & \qquad \otimes 
\left\{V_{\ul{m}}(w)+\frac{\pa V_{\ul{m}}(w)}{\pa w}(z-w) \right\}
V_{\ul{n}}(w)
\nonumber\\
 & \sim &
\frac{1}{z-w}[X,Y](w)\otimes V_{\ul{m}+\ul{n}}(w)
+ \frac{c}{(z-w)^2}(X|Y)\otimes V_{\ul{m}+\ul{n}}(w)
\nonumber\\
& & \qquad + \sum_{k=1}^M 
   \frac{m_k c}{z-w}(X|Y)\varphi^{(k)}(w)V_{\ul{m}+\ul{n}}(w). 
\end{eqnarray}
Comparing the last line to \eqref{torOPE}, we have the desirous result.
\end{proof}

\noindent{\bf Remark:} 
In the preceding works \cite{BB,IT,ISW1,ISW2}, a much bigger Lie algebra
that includes the derivations to $\gtor$ is considered. Here we do not
consider the derivations since those are not needed for our purpose,
i.e., treating the $(2+1)$-dimensional NLS hierarchy. 

\bigskip

Hereafter we consider only the $\slt_2$-case to treat the $(2+1)$-dimensional 
NLS hierarchy. 
The generators of $\fraksl_2$ is denoted by $E$, $F$ and $H$ as usual: 
\begin{equation}
 [E, F] =H, \quad [H, E]=2E, \quad [H, F]=-2F. 
\end{equation}

We prepare the language of the {\it $2$-component free
fermions} \cite{JM}. Note that the notation we use below is that of
\cite{JM} and slightly different from that of \cite{IT,JMD}. 
Let ${\cal A}$ be the associative $\IC$-algebra generated by 
$\psi_j^{(\alpha)}$, $\psi_j^{(\alpha)*}$ ($j\in\IZ$, $\alpha=1,2$) 
with the relations, 
\begin{equation}
\label{CaCR}
[ \psi^{(\alpha)}_i, \psi^{(\beta)*}_j ]_+ =
\delta_{ij}\delta_{\alpha\beta}, \quad
[ \psi^{(\alpha)}_i, \psi^{(\beta)}_j ]_+ =
[ \psi^{(\alpha)*}_i, \psi^{(\beta)*}_j ]_+ = 0 .
\end{equation}
In terms of the generating series defined as
\begin{equation}
\psi^{(\alpha)}(\lambda)=\sum_{n\in\IZ}\psi^{(\alpha)}_n\lambda^n, \quad
\psi^{(\alpha)*}(\lambda)=\sum_{n\in\IZ}\psi^{(\alpha)*}_n\lambda^{-n}
\quad (\alpha=1,2), 
\end{equation}
the relation \eqref{CaCR} are rewritten as
\begin{equation}
\label{CaCRlambda}
\begin{aligned}
\left[\psi^{(\alpha)}(\lambda), \psi^{(\beta)*}(\mu)\right]_+
&= \delta_{\alpha\beta}\delta(\lambda/\mu), \\
\left[\psi^{(\alpha)}(\lambda), \psi^{(\beta)}(\mu)\right]_+
&= \left[\psi^{(\alpha)*}(\lambda), \psi^{(\beta)*}(\mu)\right]_+=0, 
\end{aligned}
\end{equation}
where $\delta(\lambda)\defeq\sum_{n\in\IZ}\lambda^n$ is the 
formal delta-function.

Consider a left ${\cal A}$-module with a cyclic vector $\vac$ satisfying
\begin{equation}
\psi_j^{(\alpha)}\vac = 0\quad (j<0), \qquad
\psi_j^{(\alpha)*}\vac =0\quad (j\geq 0) .
\end{equation}
This ${\cal A}$-module ${\cal A}\vac$ is called the fermionic Fock space,
which we denote by ${\cal F}$. 
We also consider a right ${\cal A}$-module (the dual Fock space 
${\cal F^*}$) with a cyclic vector $\dvac$ satisfying 
\begin{equation}
\dvac\psi_j^{(\alpha)} = 0\quad (j\geq 0), \qquad
\dvac\psi_j^{(\alpha)*} =0\quad (j<0). 
\end{equation}
We further define the {\it generalized vacuum vectors} as
\begin{equation}
|s_2,s_1\rangle \defeq \Psi_{s_2}^{(2)}\Psi_{s_1}^{(1)}\vac, \qquad
\langle s_1,s_2| \defeq \dvac\Psi_{s_1}^{(1)*}\Psi_{s_2}^{(2)*}, 
\end{equation}
\begin{equation}
\Psi_s^{(\alpha)} \defeq 
\begin{cases}
\psi_s^{(\alpha)*}\cdots\psi_{-1}^{(\alpha)*} & (s<0),\\
1 & (s=0),\\
\psi_{s-1}^{(\alpha)}\cdots\psi_0^{(\alpha)} & (s>0),
\end{cases}
\qquad
\Psi_s^{(\alpha)*} \defeq 
\begin{cases}
\psi_{-1}^{(\alpha)}\cdots\psi_s^{(\alpha)} & (s<0),\\
1 & (s=0),\\
\psi_0^{(\alpha)*}\cdots\psi_{s-1}^{(\alpha)*} & (s>0).
\end{cases}
\end{equation}

There exists a unique linear map (the {\it vacuum expectation value}),
\begin{equation}
{\cal F^*}\otimes_{\cal A}{\cal F} \longrightarrow \IC
\end{equation}
such that $\dvac \otimes \vac\mapsto 1.$
For $a\in{\cal A}$ we denote by
$\dvac a\vac$ the vacuum expectation value of
the vector $\dvac a\otimes \vac (=\dvac \otimes a\vac)$
in ${\cal F^*}\otimes_{\cal A}{\cal F}.$
Using the expectation value, we prepare another important notion of the
{\it normal ordering}: 
$\,:\!\psi^{(\alpha)}_i\psi^{(\beta)*}_j\!:\,\defeq 
\psi^{(\alpha)}_i\psi^{(\beta)*}_j-
\dvac \psi^{(\alpha)}_i\psi^{(\beta)*}_j\vac$. 

\begin{lemma} (\cite{DJKM,JM,JMD})
\label{lem:affine-sl2}
The operators 
\begin{equation}
\left\{
\begin{aligned}
E(z) &= \psi^{(1)}(z)\psi^{(2)*}(z), \\
F(z) &= \psi^{(2)}(z)\psi^{(1)*}(z), \\
H(z) &= \,:\!\psi^{(1)}(z)\psi^{(1)*}(z)
     \!:\,-\,:\!\psi^{(2)}(z)\psi^{(2)*}(z)\!:\,, 
\end{aligned}
\right.
\end{equation}
satisfy the OPE \eqref{affineOPE} with $c=1$, 
i.e., give a representation of $\slh_2$ on the fermionic Fock space
 ${\cal F}$. 
\end{lemma}

{}From Lemma \ref{lem:affine-sl2} and Proposition \ref{prop:tor_ext},
we have a representation of $\gtor$ on $\Ftor{y}\defeq{\cal F}\otimes
F_y$. We will use this representation in what follows to derive 
bilinear identities. 
Note that the operators $E(z)$, $F(z)$ and $H(z)$ are invariant 
under the following automorphism of fermions: 
\begin{equation}
 \label{autom}
 \iota(\psi^{(a)}_j)=\psi^{(a)}_{j+1}, \quad
 \iota(\psi^{(a)*}_j)=\psi^{(a)*}_{j+1}\qquad (j\in\IZ,\;a=1,2). 
\end{equation}

\subsection{Derivation of the bilinear identity from representation theory}
We first introduce the following operator acting on 
$\Ftor{y}\otimes\Ftor{y'}$:
\begin{equation}
\OmegaTor \defeq 
\sum_{\ul{m}\in\IZ^M}\sum_{\alpha=1,2}\oint\frac{\dd\lambda}{2\pi\ii\lambda}
\psi^{(\alpha)}(\lambda)V_{\ul{m}}(\ul{y};\lambda)\otimes
\psi^{(\alpha)*}(\lambda)V_{-\ul{m}}(\ul{y}';\lambda) . 
\end{equation}
\begin{lemma}
\label{Lem:PropofOmegaTor}
The operator $\OmegaTor$ enjoys the following properties: 
\begin{eqnarray}
\mbox{(i)} && [ \OmegaTor,\, \slt_2\otimes 1 + 1\otimes\slt_2] = 0, \\
\mbox{(ii)} && \OmegaTor \left( |s_2,s_1\rangle\otimes 1\right)^{\otimes 2}=0. 
\label{OmegaVac1}
\end{eqnarray}
\end{lemma}
\begin{proof}
Since the representation of $\slt_2$ under consideration is constructed 
from Lemma \ref{lem:affine-sl2}, it is enough to show 
\begin{equation}
\left[ \OmegaTor,\, 
\psi^{(\alpha)}(p)\psi^{(\beta)*}(p)V_{\ul{n}}(\ul{y};p)\otimes 1
+ 1\otimes\psi^{(\alpha)}(p)\psi^{(\beta)*}(p)V_{\ul{n}}(\ul{y}';p)
\right] = 0, 
\end{equation}
for $\alpha,\beta=1,2$ and $\ul{n}\in\IZ^M$. From \eqref{CaCRlambda}, we have 
\begin{equation}
\begin{aligned}
\left[\psi^{(\alpha)}(p)\psi^{(\beta)*}(q), \psi^{(\gamma)}(\lambda)\right]
 &= \delta_{\beta\gamma}\delta(q/\lambda)\psi^{(\alpha)}(p), \\
\left[\psi^{(\alpha)}(p)\psi^{(\beta)*}(q), \psi^{(\gamma)*}(\lambda)\right]
 &= -\delta_{\alpha\gamma}\delta(p/\lambda)\psi^{(\beta)}(q). 
\end{aligned} 
\end{equation}
These equations and the relation $V_{\ul{m}}(\ul{y};\lambda)
V_{\ul{n}}(\ul{y};\lambda)=V_{\ul{m}+\ul{n}}(\ul{y};\lambda)$ 
give the commutativity above. 
\end{proof}

If we translate Lemma \ref{Lem:PropofOmegaTor} into bosonic language,
then it comes out a hierarchy of Hirota bilinear equations. 
To do this, we present a summary of the {\it boson-fermion correspondence} 
in the $2$-component case. 
Define the operators $H_n^{(\alpha)}$ as 
$H_n^{(\alpha)} \defeq \sum_{j\in\IZ}
\psi^{(\alpha)}_j \psi^{(\alpha)*}_{j+n}$
for $n=1,2,\dots$, $\alpha=1,2$, which obey the canonical commutation relation 
$
[ H^{(\alpha)}_m, H^{(\beta)}_n ] = 
m\delta_{m+n,0}\delta_{\alpha\beta}\cdot 1.
$
The operators $H_n^{(\alpha)}$ generate the Heisenberg subalgebra 
({\it free bosons}) of ${\cal A}$, which is isomorphic to the algebra 
with the basis $\{ nx^{(\alpha)}_n,\, \pa/\pa x^{(\alpha)}_n \:
(\alpha=1,2,\, n=1,2,\dots)\}$. 
\begin{lemma} (\cite{DJKM,JM,JMD})
\label{Lem:BF}
For any $|\nu\rangle\in{\cal F}$ and $s_1,s_2\in\IZ$, 
we have the following formulas, 
\begin{eqnarray}
\lefteqn{\langle s_1,s_2|\ee^{H(\ul{x}^{(1)},\ul{x}^{(2)})}
\psi^{(1)}(\lambda)|\nu\rangle}\qquad\nonumber\\
&=& (-)^{s_2}\lambda^{s_1-1}\ee^{\xi(\ul{x}^{(1)},\lambda)}
\langle s_1-1,s_2|
\ee^{H(\ul{x}^{(1)}-[\lambda^{-1}],\ul{x}^{(2)})}|\nu\rangle, \\
\lefteqn{\langle s_1,s_2|\ee^{H(\ul{x}^{(1)},\ul{x}^{(2)})}
\psi^{(1)*}(\lambda)|\nu\rangle}\qquad\nonumber\\
&=& (-)^{s_2}\lambda^{-s_1}\ee^{-\xi(\ul{x}^{(1)},\lambda)}
\langle s_1+1,s_2|
\ee^{H(\ul{x}^{(1)}+[\lambda^{-1}],\ul{x}^{(2)})}|\nu\rangle, \\
\lefteqn{\langle s_1,s_2|\ee^{H(\ul{x}^{(1)},\ul{x}^{(2)})}
\psi^{(2)}(\lambda)|\nu\rangle}\qquad\nonumber\\
&=& \lambda^{s_2-1}\ee^{\xi(\ul{x}^{(2)},\lambda)}
\langle s_1,s_2-1|
\ee^{H(\ul{x}^{(1)},\ul{x}^{(2)}-[\lambda^{-1}])}|\nu\rangle, \\
\lefteqn{\langle s_1,s_2|\ee^{H(\ul{x}^{(1)},\ul{x}^{(2)})}
\psi^{(2)*}(\lambda)|\nu\rangle}\qquad\nonumber\\
&=& \lambda^{-s_2}\ee^{-\xi(\ul{x}^{(2)},\lambda)}
\langle s_1,s_2+1|
\ee^{H(\ul{x}^{(1)},\ul{x}^{(2)}+[\lambda^{-1}])}|\nu\rangle, 
\end{eqnarray}
where the ``Hamiltonian'' $H(\ul{x}^{(1)},\ul{x}^{(2)})$ is defined as 
\begin{equation}
H(\ul{x}^{(1)},\ul{x}^{(2)}) \defeq \sum_{\alpha=1,2}
\sum_{n=1}^{\infty}x_n^{(\alpha)}H_n^{(\alpha)}. 
\end{equation}
\end{lemma}
We prepare one more lemma due to Billig \cite{Bi}. 
\begin{lemma} (\cite{Bi}, Proposition 3. See also \cite{ISW2}) 
\label{Lem:Billig}
Let $P(n)=\sum_{j\geq 0}n^j P_j$, where $P_j$ are differential operators 
that may not depend on $z$. If 
$
 \sum_{n\in\IZ}z^n P(n)f(z)=0
$
for some function $f(z)$, then 
$
 \left.P(\epsilon-z\pa_z)f(z)\right|_{z=1}=0
$
as a polynomial in $\epsilon$.
\end{lemma}

Now we are in position to state the bilinear identity for the 
$(2+1)$-dimensional NLS hierarchy. 
Let $\SL_2^{\mathrm{tor}}$ denote a group of invertible linear 
transformations on $\Ftor{y}$ generated by the exponential action of the
elements in $\fraksl_2\otimes R$ acting locally nilpotently. 
Define the $\tau$-function associated with
$\bm{g}\in\SL_2^{\mathrm{tor}}$ as 
\begin{equation}
\label{DefOfTau}
\tau^{s'_1,s'_2}_{s_2,s_1}(\ul{x}^{(1)},\ul{x}^{(2)},\ul{y})\,\defeq \,
^{\mathrm{tor}}\langle s'_1,s'_2|\ee^{H(\ul{x}^{(1)},\ul{x}^{(2)})}
\bm{g}(\ul{y})
|s_2,s_1\rangle^{\mathrm{tor}}, 
\end{equation}
where $|s_2,s_1\rangle^{\mathrm{tor}}\defeq |s_2,s_1\rangle\otimes 1$
and 
$^{\mathrm{tor}}\langle s'_1,s'_2|\defeq\langle s'_1,s'_2|\otimes 1$. 
Hereafter we shall omit the superscripts ``tor'' 
if it does not cause confusion. 
Since $\bm{g}\in\SL_2^{\mathrm{tor}}$, the $\tau$-function \eqref{DefOfTau}
have the following properties \cite{JM}: 
\begin{align}
& \tau^{s_1+\ell+1,s_2-\ell+1}_{s_2+1,s_1+1}
= (-1)^{\ell}\tau^{s_1+\ell,s_2-\ell}_{s_2,s_1}, \label{sl2}\\
& \left(\frac{\pa}{\pa x_j^{(1)}}+\frac{\pa}{\pa x_j^{(2)}}\right)
\tau^{s'_1,s'_2}_{s_2,s_1}=0, 
\end{align}
i.e., the $\tau$-function
depends only on $\{x_j\defeq x_j^{(1)}-x_j^{(2)}\}$ and $\{y_j\}$. 
\begin{proposition}
\label{Thm:BI1}
For non-negative integers $k$, $l_1$ and $l_2$, the $\tau$-functions 
satisfy 
\begin{align}
(-1)&^{s'_2+s''_2} \oint \frac{\dd\lambda}{2\pi\ii}
\lambda^{s'_1-s''_1+k-2}
\ee^{\xi((\ul{x}-\ul{x}')/2,\lambda)}\nonumber\\
&\times \tau^{s'_1-1,s'_2}_{s_2+l_2,s_1+l_1}
(\ul{x}-[\lambda^{-1}],\ul{y}-\ul{b}_{\lambda})
\tau^{s''_1+1,s''_2}_{s_2,s_1}
(\ul{x}'+[\lambda^{-1}],\ul{y}+\ul{b}_{\lambda})
\nonumber\\
+\; & \oint \frac{\dd\lambda}{2\pi\ii}
\lambda^{s'_2-s''_2+k-2}\ee^{\xi((\ul{x}'-\ul{x})/2,\lambda)}
\nonumber\\
&\times\tau^{s'_1,s'_2-1}_{s_2+l_2,s_1+l_1} 
(\ul{x}+[\lambda^{-1}],\ul{y}-\ul{b}_{\lambda})
\tau^{s''_1,s''_2+1}_{s_2,s_1}
(\ul{x}'-[\lambda^{-1}],\ul{y}+\ul{b}_{\lambda})
= 0 .\label{2cBI1}
\end{align}
\end{proposition}
\begin{proof}
This is the direct consequence of Lemmas 
\ref{Lem:PropofOmegaTor}, \ref{Lem:BF}, \ref{Lem:Billig}.
\end{proof}
Setting 
\begin{equation}
\label{tautoFGGtil}
 F=\tau^{0,0}_{0,0}, \qquad G=\tau^{1,-1}_{0,0}, \qquad
 \tilde{G}=-\tau^{-1,1}_{0,0}, 
\end{equation}
one can show that \eqref{2cBI1} contains the bilinear equations 
\eqref{BI_FFGGtil}--\eqref{BI_GtilF}
with the condition $\check{b}+\check{c}=0$. 

We now turn to the 2-dimensional derivative NLS (DNLS) equation \cite{St1}, 
\begin{equation}
\label{2dDNLS}
\ii u_T + u_{XY} + 2\ii\left(u\int^X (|u|^2)_Y \dd X\right)_X =0. 
\end{equation}
This equation can also be treated in terms of the bilinear formulation
\cite{SOM}. Following Sasa et al., we set 
\begin{equation}
\label{bilinearizationDNLS}
u = \frac{fg}{\tilde{f}^2}, \qquad 
\overline{u} = -\frac{\tilde{f}\tilde{g}}{f^2}, 
\end{equation}
where we have assumed 
\begin{equation}
\label{realityDNLS}
\overline{f}=\tilde{f}, \qquad \overline{g}=-\tilde{g}. 
\end{equation}
The validity of this assumption will be
discussed in the next section. 
These $u$ and $\overline{u}$ solve \eqref{2dDNLS} if the variables $f$
and $g$ obey the Hirota equations, 
\begin{gather}
(\ii D_{X}D_{Y}-D_{T})f\cdot g = 0,\\
(\ii D_{X}D_{Y}+D_{T})\tilde{g}\cdot\tilde{f} = 0,\\
(\ii D_{X}D_{Y}+2D_{T})f\cdot\tilde{f} = D_{Y}\tilde{g}\cdot g,
\label{BilinDNLS3}\\
\ii D_X f \cdot\tilde{f} = g\tilde{g}. \label{BilinDNLS4}
\end{gather}
We note that our bilinearization is slightly different from that of
Sasa et al. 
The first two equations can be obtained from \eqref{2cBI1} 
by making the change of the variables, 
\begin{gather}
X=\ii x_1, \quad Y=y_0, \quad T=y_1,\nonumber\\
f=\tau^{1,0}_{0,1}, \quad
g=\tau^{0,1}_{0,1}, \quad
\tilde{f}=\tau^{0,0}_{0,0}, \quad
\tilde{g}=\tau^{1,-1}_{0,0}. 
\label{fgtilftilg}
\end{gather}

The bilinear identity including the rest two can be obtained
in the same way as Proposition \ref{Thm:BI1}: 
\begin{proposition}
\label{Thm:BI2}
For non-negative integers $k$, the $\tau$-functions 
satisfy 
\begin{align}
(-1)&^{s'_2+s''_2} \oint \frac{\dd\lambda}{2\pi\ii}
\lambda^{s'_1-s''_1+k-2}
\ee^{\xi((\ul{x}-\ul{x}')/2,\lambda)}\nonumber\\
&\times \tau^{s'_1-1,s'_2}_{s_2,s_1}
(\ul{x}-[\lambda^{-1}],\ul{y}-\ul{b}_{\lambda})
\tau^{s''_1+1,s''_2}_{s_2,s_1+1}
(\ul{x}'+[\lambda^{-1}],\ul{y}+\ul{b}_{\lambda})
\nonumber\\
+\; & \oint \frac{\dd\lambda}{2\pi\ii}
\lambda^{s'_2-s''_2+k-2}\ee^{\xi((\ul{x}'-\ul{x})/2,\lambda)}
\nonumber\\
&\times\tau^{s'_1,s'_2-1}_{s_2,s_1} 
(\ul{x}+[\lambda^{-1}],\ul{y}-\ul{b}_{\lambda})
\tau^{s''_1,s''_2+1}_{s_2,s_1+1}
(\ul{x}'-[\lambda^{-1}],\ul{y}+\ul{b}_{\lambda})\nonumber\\
=\; & \tau^{s'_1,s'_2}_{s_2,s_1+1} (\ul{x},y_0,\check{y}-\check{b})
  \tau^{s''_1,s''_2}_{s_2,s_1}(\ul{x}',y_0,\check{y}+\check{b})
 .\label{2cBI2}
\end{align}
\end{proposition}
\begin{proof}
Using 
\begin{equation}
\OmegaTor \left( 
|s_2,s_1\rangle^{\mathrm{tor}}\otimes 
|s_2,s_1+1\rangle^{\mathrm{tor}}\right)=
\left(|s_2,s_1+1\rangle\otimes\ee^{my_0}\right)
\otimes\left(|s_2,s_1\rangle\otimes\ee^{-my'_0}\right)
\end{equation}
instead of \eqref{OmegaVac1}, we can derive the desirous result.
\end{proof}
Expanding \eqref{2cBI2}, we can obtain
Hirota-type differential equations including the following ones:   
\begin{gather}
(D_{x_1}D_{y_0}-2D_{y_1})
\tau^{1,0}_{0,1}\cdot\tau^{0,0}_{0,0}
= D_{y_0}\tau^{1,-1}_{0,0}\cdot\tau^{0,1}_{0,1}, \\
D_{x_1}\tau^{1,0}_{0,1}\cdot\tau^{0,0}_{0,0}
+ \tau^{0,1}_{0,1}\tau^{1,-1}_{0,0}= 0 . 
\end{gather}
These equations agree with \eqref{BilinDNLS3} and \eqref{BilinDNLS4}. 

\subsection{Reality conditions and soliton-type solutions}
In this section, we consider an algebraic meaning of the reality
condition \eqref{reality}. To this aim, we introduce an automorphism
$\rho$ of the fermion algebra as 
\begin{equation}
\rho(\psi_{n}^{(\alpha)}) = \psi_{-n-1}^{(\alpha)*}, \quad 
\rho(\psi_{n}^{(\alpha)*}) = \psi_{-n-1}^{(\alpha)} \quad (n\in\IZ,\, \alpha=1,2), 
\end{equation}
which have the following properties: 
\begin{itemize}
 \item $\rho^2=\mathrm{id}$, 
 \item $\rho(H_n^{(\alpha)})=-H_n^{(\alpha)}$\quad ($\alpha=1,2$),
 \item $\dvac\rho(\bm{g})\vac =\dvac\bm{g}\vac$
         ,\quad $^{\forall}\bm{g}\in\SL_2^{\mathrm{tor}}$. 
\end{itemize}
We note that the similar automorphism have been discussed by
Jaulent, Manna and Martinez-Alonso \cite{JMM}. 
Assuming the conditions 
\begin{equation}
\label{realNLS}
x^{(i)}_n\in\ii\IR \;\; (n\in\IN,\, i=1,2), \quad
\rho(\bm{g})=\overline{\bm{g}}, 
\end{equation}
we find that 
\begin{align}
\overline{
\langle 0,0|\ee^{H(\ul{x}^{(1)},\ul{x}^{(2)})}\bm{g}|0,0\rangle}
& =\langle 0,0|\ee^{H(\ul{x}^{(1)},\ul{x}^{(2)})}
\bm{g}|0,0\rangle, \\
\overline{\langle 1,-1|\ee^{H(\ul{x}^{(1)},\ul{x}^{(2)})}
\bm{g}|0,0\rangle}
& =\langle -1,1|
\ee^{H(\ul{x}^{(1)},\ul{x}^{(2)})}\bm{g}|0,0\rangle . 
\end{align}
Under these conditions, the $\tau$-functions $F$, $G$, $\tilde{G}$ of
\eqref{tautoFGGtil} satisfy the reality condition, 
\begin{equation}
 \overline{F}=F, \qquad \overline{G}=-\tilde{G}. 
\end{equation}

Next we introduce another automorphism $\sigma$ to treat the
$(2+1)$-dimensional DNLS equation \eqref{2dDNLS}: 
\begin{equation}
\begin{aligned}
\sigma( \psi^{(1)}_n ) = \psi^{(2)}_n , \quad 
&\sigma( \psi^{(1)*}_n ) = \psi^{(2)*}_n , \\
\sigma( \psi^{(2)}_n ) = \psi^{(1)}_{n+1}, \quad
&\sigma( \psi^{(2)*}_n ) = \psi^{(1)*}_{n+1}, 
\end{aligned}
\end{equation} 
which have the following properties, 
\begin{itemize}
 \item If $\;\iota(\bm{g})=\bm{g}$, \ then $\;\sigma^2(\bm{g})=\bm{g}$,
 \item $\sigma(H_n^{(1)})=H_n^{(2)}$, \quad $\sigma(H_n^{(2)})=H_n^{(1)}$, 
 \item $\langle 1,0|\sigma(\bm{g})|0,1\rangle
        =\dvac\bm{g}\vac$
        ,\quad $^{\forall}\bm{g}\in\SL_2^{\mathrm{tor}}$. 
\end{itemize}
Imposing the conditions 
\begin{equation}
\label{realDNLS}
x^{(1)}_n = \overline{x^{(2)}_n}\;\;(n\in\IN), \quad
\sigma(\bm{g})=\overline{\bm{g}}, 
\end{equation}
we find that 
\begin{align}
\overline{\langle 0,0|\ee^{H(\ul{x}^{(1)},\ul{x}^{(2)})}
\bm{g}|0,0\rangle}
& =\langle 1,0|\ee^{H(\ul{x}^{(1)},\ul{x}^{(2)})}
\bm{g}|0,1\rangle , \\
\overline{\langle 1,-1|\ee^{H(\ul{x}^{(1)},\ul{x}^{(2)})}
\bm{g}|0,0\rangle}
& = -\langle 0,1|\ee^{H(\ul{x}^{(1)},\ul{x}^{(2)})}
\bm{g}|0,1\rangle . 
\end{align}
In this case, the $\tau$-functions $f$, $g$, $\tilde{f}$, $\tilde{g}$
of \eqref{fgtilftilg} satisfy the reality condition
\eqref{realityDNLS}. 

As an example of special solutions, we consider soliton-type solutions
given by 
\begin{gather}
\tau^{s'_1,s'_2}_{s_2,s_1}(\ul{x}^{(1)},\ul{x}^{(2)})
= \langle s'_1,s'_2|\ee^{H(\ul{x}^{(1)},\ul{x}^{(2)})}
\bm{g}_N(\ul{y})|s_2,s_1\rangle ,  \\
\begin{array}{rl}
\displaystyle\bm{g}_N(\ul{y}) \defeq \prod_{j=1}^N 
\lefteqn{\exp\! \left[ 
a_j\psi^{(1)}(p_j)\psi^{(2)*}(p_j)V_{m_j}(\ul{y};p_j)\right.}
& \\
& \qquad \left. +b_j\psi^{(2)}(q_j)\psi^{(1)*}(q_j)V_{n_j}(\ul{y};q_j)
\right] . 
\end{array}
\end{gather}
In the NLS case $\rho(\bm{g}_N)=\overline{\bm{g}_N}$, the parameters should obey the
conditions, 
\begin{equation}
 q_j = \overline{p_j}, \quad b_j=-\overline{a_j} \quad (j=1,\dots,N). 
\end{equation}
In the DNLS case $\sigma(\bm{g}_N)=\overline{\bm{g}_N}$, we have 
\begin{equation}
 q_j = \overline{p_j}, \quad b_j=\overline{a_j p_j} \quad (j=1,\dots,N). 
\end{equation}

We conclude that the NLS case and the DNLS case have different complex
structure that correspond to different real forms of the toroidal Lie
algebra $\slt_2$. 

\subsection{Bilinear identity for the SDYM hierarchy}
The SDYM equation can also be treated also by the Hirota's bilinear
method \cite{SOM}. Toward this aim, we shall take so-called ``Yang's
$R$-gauge'' defined as follows: 
Due to \eqref{SDYM1}, there exist matrix-valued functions $\bm{G}$ and
$\bar{\bm{G}}$ such that 
\begin{equation}
\left\{\begin{aligned}
\pa_y \bm{G} &=& \bm{G}\bm{A}_y, \\
\pa_z \bm{G} &=& \bm{G}\bm{A}_z,
\end{aligned}\right.
\qquad
\left\{\begin{aligned}
\pa_{\by}\bar{\bm{G}} &=& \bar{\bm{G}}\bm{A}_{\by}, \\
\pa_{\bz}\bar{\bm{G}} &=& \bar{\bm{G}}\bm{A}_{\bz}. 
\end{aligned}\right.
\end{equation}
If we define the matrix $\bm{J}$ as 
$\bm{J}\defeq \bm{G}\bar{\bm{G}}^{-1}$, 
the self-duality equation \eqref{SDYM1} takes the form
\begin{equation}
\label{SDYM2}
\pa_{\by} \left( \bm{J}^{-1}\pa_{y}\bm{J} \right) + 
\pa_{\bz} \left( \bm{J}^{-1}\pa_{z}\bm{J} \right) =0. 
\end{equation}
We then consider the gauge field $\bm{J}$ of the form, 
\begin{equation}
\label{JtoTau}
\bm{J}=\frac{1}{f}
\begin{pmatrix}
1 & -g\\ e & f^2-eg
\end{pmatrix}
, \quad
e=\frac{\tau_1}{\tau_5},\quad
f=\frac{\tau_2}{\tau_5},\quad
g=\frac{\tau_3}{\tau_5}. 
\end{equation}
The gauge field $\bm{J}$ of \eqref{JtoTau} solves \eqref{SDYM2} if the
$\tau$-functions satisfy the following seven Hirota-type equations
\cite{SOM}, 
\begin{gather}
\tau_5^2+\tau_2\tau_8-\tau_4\tau_6=0,
\label{BilinSDYM1}\\
D_y\tau_1\cdot\tau_5 = D_{\bar{z}}\tau_4\cdot\tau_2,
\label{BilinSDYM2a}\\
D_y\tau_2\cdot\tau_6 = D_{\bar{z}}\tau_5\cdot\tau_3, 
\label{BilinSDYM2b}\\
D_y\tau_4\cdot\tau_8 = D_{\bar{z}}\tau_5\cdot\tau_7, 
\label{BilinSDYM2c}\\
D_z\tau_1\cdot\tau_5 = D_{\bar{y}}\tau_2\cdot\tau_4, 
\label{BilinSDYM3a}\\
D_z\tau_2\cdot\tau_6 = D_{\bar{y}}\tau_3\cdot\tau_5, 
\label{BilinSDYM3b}\\
D_z\tau_4\cdot\tau_8 = D_{\bar{y}}\tau_7\cdot\tau_5, 
\label{BilinSDYM3c}
\end{gather}
where we have introduced auxiliary dependent variables $\tau_4$,
$\tau_6$, $\tau_7$, $\tau_8$. 

The bilinear identity associated with these equations is given as
follows: 
\begin{proposition}
\label{Thm:BI3}
For non-negative integers $k$, the $\tau$-functions 
satisfy 
\begin{align}
(-1)&^{s'_2+s''_2} \oint \frac{\dd\lambda}{2\pi\ii}
\lambda^{s'_1-s''_1+k-2}
\ee^{\xi((\ul{x}-\ul{x}')/2,\lambda)}\nonumber\\
&\times \tau^{s'_1-1,s'_2}_{s_2,s_1}
(\ul{x}-[\lambda^{-1}],\ul{y}-\ul{b}_{\lambda})
\tau^{s''_1+1,s''_2}_{s_2+1,s_1+1}
(\ul{x}'+[\lambda^{-1}],\ul{y}+\ul{b}_{\lambda})
\nonumber\\
+\; & \oint \frac{\dd\lambda}{2\pi\ii}
\lambda^{s'_2-s''_2+k-2}\ee^{\xi((\ul{x}'-\ul{x})/2,\lambda)}
\nonumber\\
&\times\tau^{s'_1,s'_2-1}_{s_2,s_1} 
(\ul{x}+[\lambda^{-1}],\ul{y}-\ul{b}_{\lambda})
\tau^{s''_1,s''_2+1}_{s_2+1,s_1+1}
(\ul{x}'-[\lambda^{-1}],\ul{y}+\ul{b}_{\lambda})\nonumber\\
=\; & \tau^{s'_1,s'_2}_{s_2+1,s_1} (\ul{x},y_0,\check{y}-\check{b})
  \tau^{s''_1,s''_2}_{s_2,s_1+1}(\ul{x}',y_0,\check{y}+\check{b})
  \nonumber\\
 & -\; \tau^{s'_1,s'_2}_{s_2,s_1+1} (\ul{x},y_0,\check{y}-\check{b})
  \tau^{s''_1,s''_2}_{s_2+1,s_1}(\ul{x}',y_0,\check{y}+\check{b})
 .\label{2cBI3}
\end{align}
\end{proposition}
\begin{proof}
This can be proved in the same fashion as Proposition \ref{Thm:BI1}; 
Use 
\begin{equation}
\begin{array}{rcl}
\lefteqn{\OmegaTor \left(|s_1,s_2\rangle^{\mathrm{tor}} \otimes 
|s_1+1,s_2+1\rangle^{\mathrm{tor}}\right)}\qquad\\
 &=& \left(|s_1+1,s_2\rangle\otimes\ee^{m y_0}\right) \otimes 
     \left(|s_1,s_2+1\rangle\otimes\ee^{-m y'_0}\right)\\
 & & -\left(|s_1,s_2+1\rangle\otimes\ee^{m y_0}\right) \otimes 
     \left(|s_1+1,s_2\rangle\otimes\ee^{-m y'_0}\right)
\end{array}
\end{equation}
instead of \eqref{OmegaVac1}. 
\end{proof}
Expanding \eqref{2cBI3} and applying \eqref{sl2}, we can obtain
the following Hirota-type equations, 
\begin{gather}
(\tau^{s_1,s_2}_{s_2,s_1})^2 + 
\tau^{s_1+1,s_2}_{s_2+1,s_1}\tau^{s_1,s_2+1}_{s_2,s_1+1}
 - \tau^{s_1+1,s_2}_{s_2,s_1+1}\tau^{s_1,s_2+1}_{s_2+1,s_1}
= 0, \\
D_{y_0}\tau^{s_1+1,s_2-1}_{s_2,s_1}\cdot\tau^{s_1,s_2}_{s_2,s_1}
= D_{y_1}\tau^{s_1+1,s_2}_{s_2,s_1+1}\cdot\tau^{s_1+1,s_2}_{s_2+1,s_1},
 \\
D_{y_0}\tau^{s_1-1,s_2+1}_{s_2,s_1}\cdot\tau^{s_1,s_2}_{s_2,s_1}
= D_{y_1}\tau^{s_1,s_2+1}_{s_2,s_1+1}\cdot\tau^{s_1,s_2+1}_{s_2+1,s_1},
\end{gather}
which agree with \eqref{BilinSDYM1}--\eqref{BilinSDYM2c} if we set 
\begin{gather}
 \bar{y}=y_0, \quad z=y_1, \nonumber\\
 \tau_1=\tau^{0,0}_{1,-1}, \quad \tau_2=\ii\tau^{0,1}_{1,0}, \quad
 \tau_3=\tau^{-1,1}_{0,0}, \quad \tau_4=\ii\tau^{1,0}_{1,0}, \\
 \tau_5=\tau^{0,0}_{0,0}, \quad \tau_6=\ii\tau^{0,1}_{0,1}, \quad
 \tau_7=\tau^{1,-1}_{0,0}, \quad \tau_8=\ii\tau^{1,0}_{0,1}. \nonumber
\end{gather}
If we introduce another set of variables $\{z_j\:(j=0,1,\ldots)\}$ that
play the same role as $\{y_j\}$ and set $\bar{z}=z_0$, $y=-z_1$, 
the corresponding $\tau$-functions solve
\eqref{BilinSDYM1}--\eqref{BilinSDYM3c} simultaneously. 
We remark that the introduction of the variables $\{z_j\}$ corresponds to
the symmetry of the $3$-toroidal Lie algebra as mentioned in Section 
\ref{subsec:DefofTor}. 

To consider the reality condition for the $SU(2)$-gauge fields, 
we introduce an anti-automorphism $\kappa$ as 
\begin{equation}
\kappa(\psi_{n}^{(\alpha)}) = \psi_{n}^{(\alpha)*}, \quad 
\kappa(\psi_{n}^{(\alpha)*}) = \psi_{n}^{(\alpha)} \quad 
(n\in\IZ,\, \alpha=1,2), 
\end{equation}
which have the following properties: 
\begin{itemize}
 \item $\kappa^2=\mathrm{id}$, 
 \item $\dvac\kappa(\bm{g})\vac
         = \dvac\bm{g}\vac$,\quad 
       $^{\forall}\bm{g}\in\SL_2^{\mathrm{tor}}$. 
\end{itemize}
Using $\kappa$, 
we impose the following condition on $\bm{g}=\bm{g}(\ul{y},\ul{z})$: 
\begin{equation}
 \kappa(\bm{g}(\ul{y},\ul{z}))=\overline{\bm{g}(\ul{y},\ul{z})}. 
\end{equation}
Then we find that the $\tau$-function \eqref{DefOfTau} with 
$\ul{x}^{(1)}=\ul{x}^{(2)}=0$ obeys 
\begin{equation}
\overline{\langle s'_1,s'_2|\bm{g}(\ul{y},\ul{z})
|s_2,s_1\rangle}
= \langle s_1,s_2|\bm{g}(\ul{y},\ul{z})
|s'_2,s'_1\rangle , 
\end{equation}
and that $e$, $f$ and $g$ of \eqref{JtoTau} satisfies 
\begin{equation}
\overline{f} = -f, \qquad \overline{e} = g. 
\end{equation}
If we define $\tilde{\bm{J}}$ as 
\begin{equation}
\tilde{\bm{J}} = 
\begin{pmatrix} \omega & 0\\ 0 & \omega^{-1} \end{pmatrix} \bm{J}
\begin{pmatrix} \omega & 0\\ 0 & \omega^{-1} \end{pmatrix}, \quad
\omega = \frac{1+\ii}{\sqrt{2}}, 
\end{equation}
then $\tilde{\bm{J}}$ satisfies \eqref{SDYM2} and the reality condition 
$\overline{\tilde{\bm{J}}}=\,^t\tilde{\bm{J}}$ (See, for example, \cite{Pra}). 

\section{Concluding remarks}
We have described the hierarchy structure associated with the
$(2+1)$-dimensional NLS equation \eqref{2dNLS} based on the theory of the
KP hierarchy, and discussed several methods to construct special
solutions. Using the language of the free fermions, we have obtained the
bilinear identities from the representation of the toroidal Lie
algebras. 

The solutions constructed explicitly in this paper are limited
in the class of soliton-type.  
In case of the hierarchy of the $(2+1)$-dimensional KdV equation
\eqref{2dKdV}, 
an algebro-geometric construction of the Baker-Akhiezer function 
is indeed possible \cite{IT2}.
It may be also possible to discuss algebro-geometric (``finite-band'') 
solutions for the $(2+1)$-dimensional NLS hierarchy 
by extending our construction of the soliton-type solutions. 

Furthermore, by extending our theory, 
it may be possible to consider $(2+1)$-dimensional
generalizations of other soliton equations, such as 
the sine-Gordon equation, the Toda lattice, and so on. 
We will discuss the subjects elsewhere. 

\section*{Acknowledgments}
The authors would like to thank Dr.~Yasuhiro Ohta, Dr.~Yoshihisa Saito, 
Dr.~Narimasa Sasa for their interests and discussions. 
The first author is partially supported by 
Waseda University Grant for Special Research Project 2000A-155, 
and the Grant-in-Aid for Scientific Research (No.~12740115) from 
the Ministry of Education, Culture, Sports, Science and Technology.

\end{document}